# First Fruits of the Spitzer Space Telescope: Galactic and Solar System Studies

Michael Werner,[1] Giovanni Fazio,[2] George Rieke,[3] Thomas L. Roellig,[4] and Dan M. Watson[5]

[1]*Astronomy and Physics Directorate, Jet Propulsion Laboratory, California Institute of Technology, Pasadena, California 91109; email: mwerner@sirtfweb.jpl.nasa.gov*

[2]*Harvard-Smithsonian Center for Astrophysics, 60 Garden Street, Cambridge, Massachusetts 02138; email: gfazio@cfa.harvard.edu*

[3]*Steward Observatory, University of Arizona, Tucson, Arizona 85721; email: grieke@as.arizona.edu*

[4]*Astrophysics Branch, NASA/Ames Research Center, Moffett Field, California 94035; email: thomas.l.roellig@nasa.gov*

[5]*Department of Physics and Astronomy, University of Rochester, Rochester, New York 14627; email: dmw@pas.rochester.edu*

■ **Abstract** The *Spitzer Space Telescope*, launched in August 2003, is the infrared member of NASA's Great Observatory family. *Spitzer* combines the intrinsic sensitivity of a cryogenic telescope in space with the imaging and spectroscopic power of modern infrared detector arrays. This review covers early results from *Spitzer* that have produced major advances in our understanding of our own solar system and phenomena within the Galaxy. *Spitzer* has made the first detection of light from extrasolar planets, characterized planet-forming and planetary debris disks around solar-type stars, showed that substellar objects with masses smaller than 10 $M_{Jup}$ form through the same processes as do solar-mass stars, and studied in detail the composition of cometary ejecta in our Solar System. *Spitzer's* major technical advances will pave the way for yet more powerful future instruments. *Spitzer* should operate with full capabilities well into 2009, enabling several additional cycles of discovery and follow-up.

## 1. INTRODUCTION

The launch of the National Aeronautics and Space Administration's (NASA) *Spitzer Space Telescope* (*Spitzer*) in 2003 provided the scientific community with the most powerful tool yet available for astronomical explorations between 3.6 and 160 μm. As the infrared member of NASA's family of Great Observatories, *Spitzer* has been used very successfully in multispectral studies with its companion observatories, the *Chandra X-Ray Observatory* and the *Hubble Space Telescope* (HST). *Spitzer* has also observed objects currently accessible only in the infrared, most notably detecting radiation from extrasolar planets for the first time. This review presents a selection of early results from *Spitzer* that have advanced our understanding of the solar system and phenomena within the Galaxy. We include studies of protoplanetary and planetary debris disks where *Spitzer* results—often in combination with those from other perspectives—appear to paint a fairly complete preliminary picture. In other areas, such as observations of extrasolar planets, brown dwarf spectroscopy, and Kuiper Belt object (KBO) studies, we highlight particularly provocative or exciting results from *Spitzer*. Some topics have been largely omitted, so that this review does not fully encompass the scope of *Spitzer* results already in hand; most notably, readers interested in interstellar matter should consult individual *Spitzer* papers and refer to van Dishoeck (2004) and Cesarsky & Salama (2005) for a comprehensive summary of results from the Infrared Space Observatory (ISO). This review includes primarily results submitted for publication as of January 1, 2006. Early *Spitzer* results on extragalactic science are reported in Armus (2006) and Chary, Sheth & Teplitz (2006). The history of *Spitzer* is described in Werner (2006) and Rieke (2006).

## 2. OBSERVATORY OVERVIEW

Detailed technical descriptions of *Spitzer* and its three focal plane instruments are provided in the September 2004 special issue of the *Astrophysical Journal Supplement* (Werner et al. 2004 and following papers) and also in the proceedings of the International Society for Optical Engineering (Roellig et al. 2004b and following papers). We thus limit this technical description to those features of *Spitzer* most directly related to its scientific performance. We also highlight the innovations demonstrated by *Spitzer* that might be most applicable to future missions.

### 2.1. Orbit

*Spitzer* utilizes an Earth-trailing heliocentric orbit. As seen from Earth, *Spitzer* recedes at about 0.1 AU per year. For *Spitzer,* the Earth-trailing orbit has several major advantages over near-Earth orbits. The principal advantage is the distance from Earth and its heat; this facilitates the extensive use of radiative cooling, which makes *Spitzer*'s cryo-thermal design extremely efficient. The orbit also permits excellent sky viewing and observing efficiency. *Spitzer* is constrained to point no closer than 80º toward and no further than 120º from the Sun, but even with these constraints 35% of the sky is visible at any time, and the entire sky is visible every six months. Surprisingly, the solar orbit has enabled *Spitzer* to carry out several unique science investigations. These include: (*a*) conducting in situ studies of the Earth-trailing density enhancement in the zodiacal cloud, which *Spitzer* passes directly through, and (*b*) obtaining constraints on the nature of objects responsible for background star microlensing toward the Magellanic Clouds. *Spitzer* operates autonomously in this orbit; once or twice per day the observatory reorients itself so that a fixed X-band antenna mounted on the bottom of the spacecraft points to the Earth and downlinks 12 to 24 hours of stored data via the Deep Space Network (DSN).

### 2.2. Architecture

The overall configuration of the flight system is shown in Figure 1. The spacecraft and solar panel were provided by Lockheed Martin. The telescope, cryostat, and associated shields and shells make up the Cryogenic Telescope Assembly (CTA), built by Ball Aerospace, who also built the Multiband Imaging Photometer for *Spitzer* (MIPS) and Infrared Spectrograph (IRS) instruments. The third instrument, the Infrared Array Camera (IRAC), was built at NASA's Goddard Space Flight Center.

### 2.2.1. CRYO-THERMAL SYSTEM

Most of the CTA was at room temperature at launch; only the science instrument cold assemblies and the superfluid helium vessel were cold within the cryostat vacuum shell. This allowed a much smaller vacuum pressure vessel and a smaller observatory mass than the cold launch architecture used in the Infrared Astronomy Satellite (IRAS) and ISO missions. A combination of passive radiative cooling and helium boil-off vapor cools the components of the CTA after launch, using a concept developed by F. Low and described in Lysek et al. (1995). Radiative cooling works for *Spitzer* because the solar orbit allows the spacecraft always to be oriented with the solar array pointed toward the Sun, while the Earth is so distant that its heat input is negligible. The system of reflective and emitting shells and shields, which is always shadowed by the solar array, rejects almost all the heat that leaks inward while radiating the small amount not rejected into the cold of deep space. As a result, the outer shell of the CTA achieves a temperature of 34 to 34.5 K solely by radiative cooling.

With such a cold outer shell, a small amount of helium vapor suffices to maintain the telescope at its operating temperature, which can be as low as 5.5 K. The performance of the cryo-thermal system on-orbit has exceeded expectation. Repeated measurements of the system's helium volume suggest that *Spitzer*'s cryogenic lifetime will exceed five years, extending well into 2009.

Following cryogen depletion, the telescope will warm up, but it will still be colder than the outer shell; current estimates suggest that the temperature will be <30 K. At this temperature, both the instrumental background and the detector dark current should be low enough for *Spitzer* to continue natural background-limited operations in the shortest wavelength IRAC bands (3.6 and 4.5 μm). Additional information on the measured on-orbit thermal performance of the *Spitzer* CTA can be found in Finley, Hopkins & Schweickart (2004).

### 2.2.2. OPTICS

*Spitzer* has an f/1.6 primary mirror with a diameter of 85 cm and a Ritchey-Chretien Cassegrain optical design with system f-ratio f/12. The telescope optics and metering structure are constructed entirely of beryllium so that changes in both the telescope prescription and its alignment with the focal plane are minimized as the telescope cools on-orbit. A focus mechanism was used to adjust the axial position of the secondary mirror to optimize the focus as the telescope reached its equilibrium temperature. The resulting image quality is excellent, and the *Spitzer* telescope provides diffraction-limited performance at all wavelengths greater than 5.5 μm. The image diameter (FWHM) at 5.5 μm is ~1.3 arcsec. The three instruments are confocal to within their depths of focus.

### 2.2.3. POINTING AND REACTION CONTROL

*Spitzer*'s pointing and control system includes redundant gyroscopes and star trackers for sensing the pointing position of the telescope, and reaction wheels for moving the telescope. Visible light sensors sharing the cold focal plane determine the telescope's line of sight relative to that of the warm star trackers. The thermal design of *Spitzer* and the fixed position of the Sun relative to the spacecraft in the solar orbit give *Spitzer* a very high degree of thermal stability. Thus the warm-to-cold alignment varies by less than 0.5 arcsec over timescales of days. The star tracker has proven to be very accurate, with a noise-equivalent angle of approximately 0.11 arcsec using an average of 37 tracked stars. The excellent alignment stability and star-tracker accuracy allow *Spitzer* to point the telescope boresight to within <0.45 arcsec (1-σ radial uncertainty) of the desired position. *Spitzer* achieves excellent pointing stability of ~0.03 arcsec (1-σ) for times up to 600 seconds.

### 2.2.4. INSTRUMENTS

*Spitzer's* three instruments occupy the multiple instrument chamber behind the primary mirror. They share a common focal plane, with their fields of view defined by pickoff mirrors. The instruments achieve great scientific power with uncomplicated design through the use of state-of-the-art infrared detector arrays in formats as large as 256 × 256 pixels. For broadband imaging and low spectral resolution spectroscopy, Spitzer has achieved sensitivities close to or at the levels established by the natural astrophysical backgrounds—principally the zodiacal light—encountered in Earth orbit. The only moving part in the instrument payload is a scan mirror in the MIPS.

Together, the three instruments provide imaging and photometry in eight spectral bands between 3.6 and 160 μm and spectroscopy and spectrophotometry between 5.2 and 95 μm. Table 1 summarizes the characteristics and performance of the instruments. Compared to previous space infrared missions, most notably ISO, Spitzer brings a factor of ~10–to-100 times improvement in limiting point source sensitivity over most of its wavelength band. In addition, the arrays in use on Spitzer provide 100–to-1000 times more pixels than previously available, leading to major increases in efficiency for both imaging and spectroscopy.

**Table 1** *Spitzer* Instrumentation Summary

| λ (μm) | Array Type | λ/Δλ | F.O.V. | Pixel Size (arcsec) | Sensitivity[a] (5σ in 500 sec) |
|---|---|---|---|---|---|
| IRAC: InfraRed Array Camera—P.I. Giovanni Fazio, Smithsonian Astrophysical Observatory | | | | | |
| 3.6 | InSb | 4.7 | 5.2′ × 5.2′ | 1.22 | 1.3 μJy |
| 4.5 | InSb | 4.4 | 5.2′ × 5.2′ | 1.21 | 2.7μJy |
| 5.8 | Si:As (IBC) | 4.0 | 5.2′ × 5.2′ | 1.22 | 18 μJy |
| 8.0 | Si:As (IBC) | 2.7 | 5.2′ × 5.2′ | 1.22 | 22 μJy |
| MIPS: Multiband Imaging Photometer for SIRTF—P.I.-George Rieke, University of Arizona | | | | | |
| 24 | Si:As (IBC) | 4 | 5.4′ × 5.4′ | 2.5 | 110 μJy |
| 70 wide | Ge:Ga | 3.5 | 5.25′ × 2.6′ | 9.8 | 7.2 mJy |
| 70 fine | Ge:Ga | 3.5 | 2.6′ × 1.3′ | 5.0 | 14.4 mJy |
| 55–95 | Ge:Ga | 14–24 | 0.32′ × 3.8′ | 9.8 | 200 mJy |
| 160 | Ge:Ga (stressed) | 4 | 0.53′ × 5.3′ | 16 | 24 mJy |
| IRS: Infrared Spectrograph—P.I. Jim Houck, Cornell University | | | | | |
| 5.2–14.5 | Si:As (IBC) | 60–127 | 3.6″ × 57″ | 1.8 | 400 μJy |
| 13–18.5 (peakup imaging) | Si:As (IBC) | 3 | 1′ × 1.2′ | 1.8 | 75 μJy |
| 9.9–19.6 | Si:As (IBC) | 600 | 4.7″ × 11.3″ | 2.4 | $1.5 \times 10^{-18}$ W m$^{-2}$ |
| 14–38 | Si:Sb (IBC) | 57–126 | 10.6″ × 168″ | 5.1 | 1.7 mJy |
| 18.7–37.2 | Si:Sb (IBC) | 600 | 11.1″ × 22.3″ | 4.5 | $3 \times 10^{-18}$ W m$^{-2}$ |

[a]Sensitivity numbers are indicative of *Spitzer* performance in low-background sky. Confusion is not included. Detailed performance estimates should be based on tools available at http://ssc.spitzer.caltech.edu/obs

The Spitzer Science Center (SSC) at Caltech provides instrument handbooks and detailed performance and operability information on its Web site at http://ssc.spitzer.caltech.edu. The SSC, responsible for Spitzer science operations, serves as the main interface between Spitzer and its scientific user community.

The Jet Propulsion Laboratory, California Institute of Technology, managed Spitzer development for NASA and continues as the overall managing center while carrying out mission operations with the assistance of Lockheed Martin, Denver.

## 3. STELLAR EVOLUTION

Stars comprise most of the visible matter in the universe, making understanding star formation and evolution essential to our exploration of other fundamental questions in the field of astrophysics. In this section, we report *Spitzer*'s studies of key milestones in the stellar life cycle.

Stars form deep within dense clouds of molecular gas and dust, hidden from view at optical wavelengths but accessible in the infrared. As a young star evolves and sheds the dusty cloak of its birth, a remnant of the stellar accretion disk often remains. If this disk is dense and massive enough to support the formation of planets, it is termed a "protoplanetary" disk. In well-studied nearby regions, these disks have masses $\sim$0.001–0.01 $M_\odot$ and sizes $\sim$100 AU. As the protoplanetary disk dissipates through continued evolution and, perhaps, the formation of planets, a planetary debris disk, formed by the inevitable pulverization of the macroscopic bodies formed in the protoplanetary disk, arises to take its place. Throughout this evolution the disks are inviting targets for study in the infrared because the dust they contain is heated by the star, while the gas, when present, may be warmed by accretion, viscosity, or gas-grain collisions. The study of these protoplanetary disks remains a particularly active area of Spitzer research, and the initial results presented in Section 3.1 will be greatly augmented in the next several years.

A star leaving the main sequence passes through a series of stages that require exploration in the infrared. Copious mass loss cloaks the star in an envelope of dust that characterizes the post-main sequence asymptotic giant branch and supergiant phases. From here, lower-mass stars evolve into planetary nebulae consisting of a hot and compact white dwarf primary exciting a diffuse envelope of ejecta that radiates copiously in the infrared. More massive stars may explode as Type II supernovae, initially characterized by similar infrared-bright ejecta envelopes. At later stages, supernovae excite infrared radiation from pre-existing circumstellar and interstellar material swept up by the expanding blast wave.

Objects below the $\sim$0.08 $M_\odot$ limit required to ignite and sustain nuclear burning become brown dwarfs. These objects span almost two decades in mass from that of Jupiter (0.0015 $M_\odot$, or 1 $M_{Jup}$) to $\sim$0.08 $M_\odot$. They may have temperatures $\sim$3000 K when formed but cool rapidly and spend most of their lives at much lower temperatures, again accessible to study primarily in the infrared.

Our current understanding of star and planet formation, including but also going far beyond Spitzer's contributions, is summarized in Reipurth, Jewitt & Keil (2006).

Many of the results presented in this section and throughout this review are the initial fruits of Spitzer's Legacy science programs. Readers should refer to the sidebar to gain some appreciation of the scope of these programs, which will release large amounts of data—both from Spitzer and from ancillary sources—prior to the end of 2006. Substantial data sets are also being produced by the larger Guaranteed Time Observer (GTO) programs and by Legacy-scale programs proposed in response to the first calls for General Observer (GO) proposals.

----------------------------------------------------------------

The *Spitzer* Legacy Science Program

The original *Spitzer* Legacy Science Program enabled major science observing projects early in the *Spitzer* mission, creating a substantial and coherent database of archived observations. The six legacy projects used a total of 3160 hours of *Spitzer* observing time and released all their data to the public archive, accompanied by post-pipeline data products and analysis tools for use by the scientific community. Three of these projects focus on galactic studies, and results from these projects are presented throughout this article:

The GLIMPSE project (P.I.-E. Churchwell, Wisconsin) surveyed the galactic plane between latitude $\pm 1°$ and longitude 10–65° on both sides of the galactic center. The survey used all four IRAC bands and provides data on galactic structure, high-mass star formation, and supernova remnants and infrared dark clouds throughout the inner galaxy.

The c2d project (P.I.-N. Evans, Texas) includes complete IRAC and MIPS maps of five large molecular clouds—each covering about 4 square degrees—maps of ~135 smaller cores, and photometry and spectroscopy of about 200 stars, emphasizing objects younger than 10 million years. c2d is designed to trace the early stages of stellar evolution from starless cores to the formation of planet-forming disks.

The FEPS project (P.I.-M. Meyer, Arizona) has obtained photometry and spectroscopy of more than 300 main-sequence stars aged 3 Myr to 3 Gyr to study the evolution of dust and gas in postaccretion disk systems

-----------------------------------------------------------------

## 3.1. Formation of Stars, Substellar Objects, and Protoplanetary Disks

Lada & Wilking (1984) showed that it was useful to sort young stellar objects (YSOs) of roughly solar mass and below by the spectral energy distribution (SED) of their near- and mid-infrared excesses. In their classification system, YSO classes I, II, and III correspond to the degree to which the central object is dust-embedded. Class I corresponds to especially deeply embedded objects with heavily veiled spectra and, usually, associated outflows and Herbig-Haro objects; Class II corresponds to classical T Tauri stars with ongoing accretion and substantial circumstellar disks; and Class III corresponds simply to reddened photospheres. Adams, Lada & Shu (1987) showed that the three classes probably comprise an evolutionary sequence, initially dominated in appearance by an infalling envelope and a dense, optically-thick disk, both of which gradually dissipate. Efforts to observe stages of protostellar evolution earlier than Class I identified a rarer class of object—dubbed Class 0 YSOs—with even redder SEDs, as well as resolved, spheroidal shapes at submillimeter wavelengths, and extremely faint central objects (e.g., André, Ward-Thompson & Barsony 1993, Myers & Ladd 1993). Figure 2 shows Spitzer-obtained spectra of typical examples of the YSO classes to illustrate the decrease in dustiness along the sequence from Class 0 to Class III. Lada (2005) reviews this overall scenario.

*Spitzer* observations are producing complete studies of the spatial and temporal evolution of large samples of stars and protoplanetary systems through all of these stages; the *Spitzer* samples include objects with masses below the hydrogen-burning limit that might properly be considered young sub-stellar objects. We also present early *Spitzer* results on the formation of higher mass stars, both in the Milky Way and in the nearby Large Magellanic Clouds (LMC); *Spitzer*'s sensitivity permits detailed study of individual massive protostars at the ~55 kpc distance of the LMC.

### 3.1.1. YOUNG CLUSTERS

We have long known that most stars form in clusters. However, mid-infrared observations are needed to establish the evolutionary state of the stars. *Spitzer*'s MIPS and IRAC arrays, with their 5 × 5 arcmin fields of view and high sensitivities, quickly survey large areas at mid-infrared wavelengths, detecting large numbers of YSOs, determining their SEDs, and mapping their spatial distribution.

IRAC observations show that the IRAC color-color diagram is a very powerful diagnostic tool for rapidly identifying and classifying young stars (Allen et al. 2004, Megeath et al. 2004). Figure 3, based on data from observations of four young stellar clusters, illustrates this. Investigators are beginning to use Spitzer's ability to sort and classify YSOs to explore such questions as the timescales of the various stages of, and the propagation or distribution of, star formation through a cloud. For example, in Figure 3 the number of objects with colors between those of Class II and Class III is relatively small. This may indicate that the transition between Classes II and III is short compared to the duration of Class II (Gutermuth et al. 2006), but the overlap of Class I and II suggests that the timescale for the final dissipation or settling of the envelope is comparable to the duration of the Class I phase.

Hartmann et al. (2005b) confirm the IRAC color-color classification scheme by comparing it to the independently known properties of a well-studied sample of YSOs in Taurus. They also observe that IRAC colors are in good agreement with recent improved disk models, and in general agreement with the models for protostellar envelopes. Among the systems with disks, they observe a strong correlation between IRAC excess emission and signatures of accretion as inferred from emission line profiles or UV excess. Young et al. (2005) report MIPS observations of the Chameleon II molecular cloud and make a first effort to extend this classification into the MIPS bands. They show a [24] versus [24]–[70] color magnitude plot and identify ~16 objects brighter than fifth magnitude (~50 mJy) at 24 μm as lying within the cloud. Within this group, the known Class 0 and Class I sources are redder than the Class II objects. However, all sources generally lie blueward of the current models by Young & Evans (2005) and by Whitney et al. (2003).

Megeath et al. (2004) report on the spatial distribution of the YSOs in the four clusters used to validate IRAC's photometric classification. The clusters are similar in distance but span a range of far-infrared luminosities, molecular gas masses, and cluster membership. The results suggest that a significant number of stars in each region form outside the dense stellar cluster identified from ground-based observations, which typically has a diameter of ~1 pc. Megeath et al. (2004) also find that the distribution of sources differs from cluster to cluster, but overall the Class I objects appear to be more spatially localized than the Class II objects, as might be expected from the evolutionary sequence proposed above. NGC 7129 illustrates the occurrence of star formation outside of a cluster core. Gutermuth et al. (2004) augment IRAC data on NGC 7129 with ground-based near-infrared data and with MIPS data on the cluster from Muzerolle et al. (2004), finding that half of the YSOs are located outside the 0.5 pc cluster core and that star formation is continuing in this halo. They also determine that approximately half of the stars in the cluster core have disks. Additionally, Muzerolle et al. (2004) identify in NGC 7129 several objects with photospheric colors out to 8 μm but excesses at 24 μm; these objects must have central clearings and are probably further examples of the transition disks discussed in detail below (cf. Figure 8).

Teixeira et al.'s (2006) MIPS/IRAC observations (Figure 4) of the massive young cluster NGC 2264 identify a set of dense filaments delineated in the submillimeter continuum, which are shown by MIPS images to be dotted with bright 24-μm sources with quasi-uniform separation. The majority of these are Class I objects. The preferred spacing between these 24-μm sources, ~20 arcsec or ~0.08 pc at the distance of NGC 2264, is remarkably close to the Jeans length for these filaments. Teixeira et al. (2006) suggest that the filaments may have formed through turbulent motions of the cloud and subsequently thermalized and fragmented into star-forming cores. Young et al. (2006), combining *Spitzer* and Magellan data, resolve the submillimeter source IRAS 12 S1 in NGC 2264—previously thought to be a single Class 0 object—into a dense cluster of embedded, low-mass YSOs. This cluster's estimated dynamical lifetime of only a few times ten thousand years highlights its youth.

### 3.1.2. LARGE AREA SURVEYS

S.T. Megeath, K.M. Flaherty, J. Hora, R. Gutermuth, L.E. Allen, et al.(submitted) have mapped the Orion A and B molecular clouds, covering an area of 5.6 square degrees, with sensitivity to detect objects below the hydrogen burning limit at an age of 1 Myr. Initial results show that approximately half of the young stars and protostars identified in this survey are found in dense clusters surrounding the two regions of recent massive star formation, NGC 2024 and the Orion Nebula, whereas the other half are found in lower density environments such as L1641. This suggests that the rate and/or density of star formation may be enhanced by the presence of OB stars in the molecular cloud.

As part of the c2d Legacy program (see sidebar) Harvey et al. (2006) analyzed a 0.89-square-degree IRAC map of the Serpens dark cloud and identified more than 240 YSO candidates. They discovered a particularly rich area of star and substellar object formation about a degree southwest of the well-studied Serpens Core. These observations also suggest that a population of infrared-excess sources exists in Serpens at least down to luminosities ~0.001 $L_\odot$. Determining the nature of these low luminosity objects will require deeper imaging and spectroscopy.

### 3.1.3. THE YOUNGEST OBJECTS

Jørgensen et al. (2005) observed the Class 0 protostar 16293-2422 with the IRS, obtaining the first detection of this archetypal object shortward of 60 μm. Their detailed modeling of the SED of this object suggests a central cavity of radius ~600 AU, which they note is comparable to the centrifugal radius of the envelope. *Spitzer* observations of L1014, a dense core previously thought to be starless, showed the presence of an embedded infrared point source that Young et al. (2004a) classified as a Class 0 object. Its luminosity of at most a few tenths $L_\odot$ makes this among the lowest luminosity protostars known and suggests the formation of a brown dwarf. Bourke et al. (2005) eliminated the possibility that this might be a background object. They used the Submillimeter Array to discover an associated compact molecular outflow, which is among the smallest known—only ~500 AU in size and $<10^{-4}\ M_\odot$ in mass. Finally, Huard et al. (2006) obtained very deep near-infrared images that showed a scattered-light nebula like those typically associated with protostars. T.L. Huard et al. (in preparation) report that about 20% of cores formerly thought to be starless contain objects of very low luminosity that may be similar to the object in L1014.

Spitzer studies of the mid-infrared spectra of Class 0 and I YSOs reveal several strong absorption features of ices, silicate features that display both emission and absorption, and emission lines of abundant low-excitation ions, atoms, and molecules. Relative intensities suggest that the emission lines arise in fairly low density gas (100-to-1000 cm⁻³), probably indicating an origin in the bipolar outflows and shocks associated with these objects (E. Furlan, M. McClure, N. Calvet, L. Hartmann, P. D'Alessio, et al., submitted) rather than in gas associated with the envelopes or disks (cf. Figure 5). The 10 and 20 µm silicate features exhibit a combination of emission and absorption that is consistent with the flattening of the envelope and disk, along with a range of inclinations with respect to the line of sight (Watson et al. 2004). Their profiles resemble in detail the absorptions produced by interstellar dust grains; the silicates must be mostly amorphous, though faint hints of crystalline silicate absorption have been noted (Ciardi et al. 2005).

The ice absorption features seen in Class I and 0 YSOs by Spitzer (Boogert et al. 2004, Noriega-Crespo et al. 2004; Figures 2 and 5) also appear in ISO mid-infrared spectra of more massive deeply embedded objects (see van Dishoeck 2004). Most prominent are the features due to the ices of water ($\lambda$=6.0 and 11 µm), HCOOH ($\lambda$=7.3 µm), methane (7.8 µm), and carbon dioxide ($\lambda$=15.2 µm), and a still unidentified feature at 6.8 µm usually associated with $NH_4^+$. The Spitzer targets include much fainter and lower-mass YSOs than previously studied, and the ice features in these objects tend to be stronger relative to the silicate features than in the more massive objects (Figure 6). Studies of the Class I YSOs in Taurus (Watson et al. 2004; see also E. Furlan, M. McClure, N. Calvet, L. Hartmann, P. D'Alessio, et al., submitted) suggest that the ice absorption arises primarily in the YSO envelopes rather than in the disks. Comparison of these feature strengths to models indicates that most of the ice-feature absorption originates in the region of the envelope outside the centrifugal radius, as its strength is much less dependent upon inclination than that of the silicate feature. Thus, these observations suggest the accumulation of ices in the solid components of protoplanetary systems, seen before they settle to the disk to begin incorporation into planetesimals. Pontoppidan et al. (2005) sound a cautionary note for the interpretation of ice absorption spectra in YSOs with a detailed analysis of the spectrum of the source CRBR 2422.8-3423 showing that much of the absorption arises in the dense foreground Ophiuchus cloud. Fortunately, the Taurus Class I YSOs shown in Figure 6 are not observed through such a foreground cloud.

3.1.4. **BROWN DWARF FORMATION** Brown dwarfs with ages of only a few million years have not cooled appreciably and form a smooth continuum in luminosity and color with the slightly more massive late M stars. It had been thought that, below a certain mass, brown dwarfs might form as companions to more massive stars rather than through the gravitational collapse sequence outlined above. However, as summarized by Luhman (2006), extensive studies from *Spitzer* indicate that the frequency of occurrence and spectral properties of disks around young brown dwarfs with masses $\lesssim 10\ M_{Jup}$ do not differ from those of disks around young stars. Thus, there is as yet no evidence that brown dwarfs form differently than stars.

For example, Luhman et al. (2005c) suggest from Spitzer data that for both the IC348 and Chameleon I star-forming clusters the fraction of substellar objects in the mass range <0.08 $M_\odot$ with evidence for disks—40% to 50% in each case—does not differ significantly from the fraction of protostars with masses 0.1 $M_\odot$ to 0.7 $M_\odot$ showing evidence for disks. The statistics on substellar objects result only from IRAC observations and would not include possible transition disks (Figure 8) such as Muzerolle et al. (2006) identified using MIPS observations in IC348. Thus, these results may be treated as lower limits on the fraction of both stellar and substellar objects having disks. Hartmann et al. (2005b) report similar results demonstrating the prevalence of disks around brown dwarfs in their survey of Taurus.

Luhman et al. (2005a,b) identified young, low-mass brown dwarfs with evidence for disks, most notably a brown dwarf in Chameleon with an estimated mass of 8 $M_{Jup}$. (The mass estimate, based on comparison with brown dwarf models, is uncertain by about a factor of two.) Even taking into account the uncertainties, the masses of these disk-bearing young brown dwarfs fall squarely into the range of the planetary companions identified around nearby stars by radial velocity measurements. The SEDs of the disks around these low-mass brown dwarfs closely resemble those of the disks around true protostars. Thus the raw materials for planet formation exist around free-floating planetary-mass bodies, and the extrasolar planets closest to Earth may orbit brown dwarfs.

3.1.5. **HIGH-MASS STAR FORMATION** As part of the GLIMPSE Legacy program (see sidebar) Churchwell et al. (2004) and Whitney et al. (2004a) used *Spitzer* observations of the giant H II region RCW 49 to identify hundreds of stars forming within the clouds surrounding RCW 49. Some of the YSOs are massive B stars and are therefore very young, suggesting their formation was triggered by stellar winds and shocks generated by the older (2–3 Myr) massive central cluster.

Allen et al. (2005) surveyed W5/AFGL 4029, S255, and S235, to gain a better understanding of the processes involved in high-mass star formation. This group investigated bright-rimmed molecular clouds, where an edge-on molecular cloud surface is externally illuminated by nearby young massive stars (see Figure 7). Using the color-color diagnostics described earlier, they determined the spatial distribution and surface densities of young stars (classes I and II) and noted that the Class I objects cluster tightly on the edge of the molecular cloud, very close to the bright rim that delineates the interface between the molecular cloud and the adjacent H II region, whereas the Class II objects are more widely dispersed. These data again support the evolutionary sequence underlying the YSO classification and suggest that either the propagation of the bright rim into the cloud or the effects of previous stellar generations triggers on-going star formation in a self-sustaining fashion. Reach et al. (2004) noted similar phenomenology in their observations of the optically dark globule IC 1396A, which has a bright rim illuminated by a nearby O star.

The Midcourse Space Experiment (MSX) satellite discovered a class of infrared dark clouds that appear in silhouette at 8 µm against the bright mid-infrared emission of the galactic plane, indicative of high extinction (Egan et al. 1998). Rathborne et al. (2005) present a MIPS 24-µm image that reveals three embedded protostars in G34.4+0.2, a ~5-arcmin-long filamentary cloud of this type. Each protostar has a luminosity between 10,000 and 30,000 $L_\odot$ and a projected main sequence mass ~10 $M_\odot$. The authors suggest that the infrared dark clouds play an important role in massive star formation throughout the Galaxy. Beuther, Sridharan & Saito (2005) studied the dark cloud IRDC 18223-3. The results indicate an extremely young massive protostellar object hidden at the center of the core that causes hints of outflow activity both in molecular line emission and in IRAC imaging.

Spitzer's ability to study high-mass star formation in the LMC and other nearby galaxies seen from a favorable external perspective provides an excellent complement to its studies of similar phenomena within the Milky Way. The individual objects identified in LMC studies (e.g., Chu et al. 2005, Jones et al. 2005) have luminosities of $\sim 10^3$ to $10^4$ $L_\odot$ and inferred masses ranging upward from $\sim 10$ $M_\odot$. Van Loon et al. (2005) present the mid-infrared spectrum of a $\sim 20$ $M_\odot$ YSO in the LMC, finding a different pattern of ice abundances from that seen in our Galaxy.

3.1.6. **OUTBURSTS AND OUTFLOWS** McNeil's Nebula, illuminated by V1647 Ori, rose in brightness by about four magnitudes at I-band over a period of a few months (Briceño et al. 2004). Serendipitous MIPS and IRAC observations of V1647 Ori postoutburst, combined with data from previous surveys, provided a rare opportunity to study eruptive events in YSOs (Muzerolle et al. 2005). These observations show that the event is most likely due to a sudden increase in the accretion luminosity of the source from $\sim 3$ to $\sim 44$ $L_\odot$. They modelled the source as a Class I object with star, disk, and envelope and suggested that outbursts such as this may contribute to the clearing of Class I envelopes.

Spitzer can study the energetic outflows that accompany pre-main sequence evolution. Figure 5 shows how the different IRAC bands distinguish polycyclic aromatic hydrocarbon (PAH)-dominated reflection nebulae from knots and filaments excited by outflows in NGC 7129 (Gutermuth et al. 2004, Morris et al. 2004). The dramatic images of the HH46-47 outflow presented by Noriega-Crespo et al. (2004) show two lobes largely unseen at visible wavelengths (Figure 5) and apparently dominated by shock-heated emission. By contrast, the spectrum of the exciting star, seen through an edge-on disk, shows deep ice absorption features.

## 3.2. Evolution of Protoplanetary Disks

For a YSO of solar luminosity, the excess emission detectable by the Spitzer instruments probes the dust distribution between a few tenths and a few tens of an astronomical unit from the star, with the inner material dominating the short wavelength emission. Thus, the spectrum or SED of the infrared radiation serves neatly as a proxy for the radial structure of the disk, and the absence of an excess at the shortest wavelengths signals exhaustion or consumption of the solid material from which terrestrial planets can form.

Hartmann (2005) reviews our understanding of disk evolution as it stood prior to the availability of Spitzer data. Protoplanetary disks appear to be very common around stars at an age of 1 Myr. These disks are rarer at 10 Myr, and no optically thick disks have been found around solar-type stars older than 30 Myr (e.g., Sicilia-Aguilar et al. 2005 and references therein). To understand how disks evolve over this 30-million-year period, Spitzer observations were designed to investigate disk evolution in clusters ranging in age from less than 1 Myr to upward of 100 million years. By exploring the structure of disks in varying age ranges, Spitzer can determine the epoch of terrestrial planet formation and study the transition of protoplanetary disks into optically thin remnants and finally into planetary debris disks.

3.2.1. **TIMESCALES FOR DISK EVOLUTION IN CLUSTERS** Combining IRAC data with ground-based near-infrared fluxes, Gutermuth et al. (2004) determined that 54±14% of stars in the core of the 1-Myr-old cluster NGC 7129 exhibit disks that increase the stellar brightness by ~50% or more over the expected photospheric level at 4.5 μm. Like the other studies described in this section, the Gutermuth et al. (2004) survey reaches stellar masses below the hydrogen burning limit. Lada et al. (2006) used IRAC and MIPS to examine the ~300 members of IC 348 to investigate the frequency and nature of the circumstellar disk population in the cluster. IC 348 has an intermediate age (2–3 Myr) and lies close enough to permit a robust disk census at the peak of the stellar initial mass function. The fraction of disk-bearing stars is 44±7%; however, only 31±5% of these stars are surrounded by robust, optically thick disks. These measurements indicate that in the 2–3 Myr since the cluster formed, 70% of the stars have lost most or all of their primordial circumstellar disks. The disk fraction peaks for stars of K6–M2 spectral types, with masses similar to the Sun. Thus, planet formation appears most favorable around the solar mass stars in this cluster. Evidence for disk evolution among young brown dwarfs was presented by Muzerolle eat al. (2006), who combined the MIPS 24-μm data with the IRAC data in IC 348 for spectral types later than M6. Two objects among the six investigated show evidence for inner disk clearing. Their model for one of the objects suggests a clearing scale of ~0.5–0.9 AU; one plausible explanation is that a planet formed very rapidly close to the central object.

Sicilia-Aguilar et al. (2005) examined two clusters (Tr 37 and NGC 7160) in the Cepheus OB2 region, aged 4 and 10 Myr, respectively, both about 900 pc from the Earth. These clusters exhibited significantly different disk fractions (48% and 4%, respectively). The younger cluster shows evidence of significant disk evolution: Spitzer observations (3.6–24 μm; Sicilia-Aguilar et al. 2006) showed that about 10% of the objects with disks in Tr 37 have photospheric fluxes at wavelengths <4.5 μm and excesses at longer wavelengths, indicating an optically thin inner disk. Similarly, Megeath et al. (2005) show that the ~5-Myr-old eta Cha association shows a larger fraction of disks evolving into the optically thin stage than has been found in younger clusters. Young et al. (2004b) used IRAC and MIPS to observe the cluster NGC 2547, which is ~30 Myr old. They derived a 3.6-μm emitting disk fraction of <7% but found that ~25% of the stars showed an excess at 24 μm, suggestive of a cool disk with a central hole. It therefore appears well-established that the warm inner disks seen by IRAC dissipate on timescales less than ~10 Myr, whereas 24-μm excesses persist significantly longer.

The nearby TW Hydrae association, with an age of 8–10 Myr, provides interesting insights into disk evolution (Uchida et al. 2004, Low et al. 2005). Four of the 24 stars studied in this grouping show 24-μm excesses at a factor of ~100 over the expected photospheres—two of these are accreting T Tauri stars and two are debris systems with unusually high values of fractional luminosity, $f$, defined as a debris system's ratio of debris disk luminosity to star luminosity. Only one of the remaining 20 stars shows any evidence for a 24-μm excess, and that excess is no more than a factor of two. This sharp division between stars with very large excesses and stars with no excesses at all—particularly in such a young population—suggests that the transitions between these states occur very quickly. Zuckerman & Song (2004) review the use of nearby clusters, such as the TW Hydrae association, for the study of disk evolution.

An exception to the rule that dust disks around newborn stars disappear in a few million years was the discovery by Spitzer of a dusty disk orbiting St 34, an accreting M star binary with an estimated age of ~25 Myr (Hartmann et al. 2005a). They suggest that the dust persists because dynamical effects of the binary star have inhibited planet formation, which otherwise would have dissipated the orbiting material.

3.2.2. **SPECTROSCOPIC STUDIES OF DISK EVOLUTION** The possible contribution of PAH emission and silicate emission or absorption in the IRAC channels introduces uncertainty in the interpretation of colors in young clusters. This is alleviated by extracting the dust features and the underlying continuum separately from IRS spectra. Furlan et al. (2005a) recently constructed the first such "emission-feature free" mid-infrared color-color diagram for a complete sample of Class II objects in the Taurus cloud. The match between the data and models is poor unless dust grains are heavily depleted by settling to the disk midplane; it appears that in most of these Class II objects, small dust grains are depleted from the upper reaches of the disks by factors of $100^{-}1000$. Settling of dust to the midplane of a protoplanetary disk due to gas drag forces may initiate the planetary formation process by stimulating grain-grain collisions and the coalescence of increasingly massive solid particles (Lada 2005).

The coalescence process also leaves an imprint on the shape and crystallinity of the 10 and 20 μm silicate emission features, as originally shown by ISO for Herbig Ae stars (van Boekel et al. 2005 and references therein). Spitzer spectroscopy has identified and studied in detail this coalescence in solar mass YSOs (Forrest et al. 2004, Kessler-Silacci et al. 2006). In addition, Apai et al. (2005) report IRS studies of young brown dwarfs that demonstrate both grain growth and the settling of grains to the midplane in the disks around these substellar objects.

Disks with central clearings are visible above the stellar photosphere only at longer wavelengths than are radially continuous disks. IRS spectra can exploit this phenomenon in detail to study the structure of the disks and the range of mechanisms responsible for the central clearings in these transition disks. A spectroscopic survey of 150 members of Taurus-Auriga (Forrest et al. 2004; also E. Furlan, L. Hartmann, N. Calvet, P. D'Alessio, R. Franco-Hernandez, et al., submitted) has so far yielded four confirmed transition disks. The spectra (Figure 8) show clearly that the inner edges of the disks are very sharp, accurately modeled in each case as a single-temperature blackbody wall with a thin atmosphere, and that the central clearings are practically devoid of small dust grains (Calvet et al. 2005).

As enumerated by D'Alessio et al. (2005a), radiative, stellar-wind, or gas-drag mechanisms fail to produce central clearings in dusty disks within the ages of the Taurus stars. However, a companion, formed after both the disk and central star, could produce a clearing through its orbital resonances that prevent the exterior disk from moving further in but allows the inner material to accrete onto the star. Simulations by Quillen et al. (2004) and Varniere et al. (2006a,b) show that central clearings such as that in the CoKu Tau/4 disk (Figure 8) can be produced by companions of the Saturn-Jupiter class in less than $10^5$ years. Establishing that massive planetary companions produce these gaps would challenge theories of planet formation, because the ages of the stars in these systems (1–2 Myr) lie in the no-man's land between the timescales for giant-planet formation in the two leading models. To form a Jovian planet by core-accretion processes (e.g., Pollack et al. 1996) takes upward of 10 Myr. To form one by growth of gravitational instabilities in a gaseous disk takes only on the order of 1000 years (e.g., Boss 2005); however, the disks may be gravitationally unstable only in the earliest evolutionary phases, and a planet formed too early may migrate inward to the star within 1 Myr.

Spitzer spectra can further constrain disk evolution by studying the gas content of disks of various ages. Lahuis et al. (2006) report the discovery of gas phase absorption due to $C_2H_2$, HCN, and $CO_2$ in the low-mass Class I YSO IRS 46 in Ophiuchus. The high rotational temperatures suggest that this material is located in the terrestrial planet zone of the disk. This is the only YSO in a sample of 100 that shows these features. This may reflect mainly the need for both favorable geometry and a large intrinsic line width to make the lines visible in absorption, as many of these ~1 Myr old objects should retain substantial amounts of their primordial gas.

In the case of HD105, with an age of ~30 Myr, Hollenbach et al. (2005) obtain upper limits on the intensity of several infrared emission lines and show that less than 1 $M_{Jup}$ of gas exists between 1 and 40 AU, so that giant-planet formation cannot occur in this disk at present; however, there may be enough residual gas in the disk to influence the dynamics of the dust.

3.2.3. **CRYSTALLINE SILICATES AMONG SMALL DUST GRAINS IN CLASS II YOUNG STELLAR OBJECTS** Evolution of silicate dust grain composition in protoplanetary disks has been studied for more than a decade, via ground-based (e.g., Knacke et al. 1993, Honda et al. 2003, Kessler-Silacci et al. 2005) and ISO spectroscopy (van Dishoeck 2004). *Spitzer* spectra of Class II YSOs in this domain show evidence of silicate compositions ranging from one indistinguishable from that of amorphous interstellar grains to one heavily dominated by crystalline minerals such as pyroxenes, olivines, and silica. In Figure 2 the spectra of FM Tau and IRAS F04147+2822 serve as examples of the extremes. The latter shows the same crystalline silicate emission features attributable to forsterite (crystalline $Mg_2SiO_4$) that appear in the spectra shown in Figure 9. ISO spectra of more massive luminous Herbig Ae stars show a similar variety (van Boekel et al. 2005).

Among Class II YSOs, Spitzer spectra indicate that the crystalline mass fraction among the small dust grains varies from very small upper limits below those placed on interstellar grains (<2%; Kemper, Vriend & Tielens 2004), to nearly 100% (Forrest et al. 2004, Uchida et al. 2004, Kessler-Silacci et al. 2006, Sargent et al. 2006; also D.M. Watson, J.D. Leisenring, E. Furlan, C.J. Bohac, B. Sargent, W.J. Forrest, et al., submitted). Not even the lowest-mass objects are immune to the large range of thermal processing implied, and several brown dwarfs and brown-dwarf candidates have displayed rich crystalline silicate spectra (Apai et al. 2005, Furlan et al. 2005a, Sargent et al. 2006; also B. Merin, et al in preparation; see Figure 9). The quantification of the degree of crystallinity can be confounded by the simultaneous evolution of the particle size distribution. In several cases, detailed modeling suggests that the Spitzer spectra indicate substantial fractions of large (>1 μm), porous dust grains (Kessler-Silacci et al. 2006, Sargent et al. 2006). There are even indications of differences in the composition of amorphous grain material among the disks of low-mass Class II objects (Sargent et al.

2006). These strong indications of grain processing correlate with the observation of dust grain growth and settling to the disk midplane (Furlan et al. 2005b; also E. Furlan, L. Hartmann, N. Calvet, P. D'Alessio, R. Franco-Hernandez, et al., submitted), which in turn implies the potential in these objects for rocky planetesimal growth and planet formation. As shown in Figure 9, Spitzer spectra of small bodies in the solar system show crystalline silicate emission very similar to that seen in protoplanetary and debris disks.

The origin of the crystalline silicates remains a mystery (see Molster & Kemper 2005). E. Furlan, L. Hartmann, N. Calvet, P. D'Alessio, R. Franco-Hernandez, et al. (submitted), Kessler-Silacci et al. (2006), and D.M. Watson, J.D. Leisenring, E. Furlan, C.J. Bohac, B. Sargent, et al. (submitted), have studied the emission spectra of large samples of Class II objects in order to understand how the amorphous interstellar silicates convert to crystalline form. They find no significant correlation between crystallinity and any property of the disks or central stars, including stellar mass, effective temperature, and accretion rate, indicating no systematic influence of radiation. There is also no correlation with disk mass or disk temperature, nor with the ratio of disk mass to stellar mass, which discounts a connection with disk-based heating mechanisms such as spiral shocks. A correlation of crystalline mass fraction with X-ray properties (flaring, in particular) remains to be explored, as does whether the structural evolution of the disks, driven by the formation and migration of planets, produces the lack of correlation between crystallinity and stellar or disk properties[1] (D.M. Watson, J.D. Leisenring, E. Furlan, C.J. Bohac, B. Sargent, et al., submitted).

### 3.3. Post-main-Sequence Stellar Evolution

Post-main-sequence stellar evolution virtually always involves mass loss and dust condensation. The resulting phenomena that can be studied with Spitzer range from strong infrared continua associated with dust, through interesting chemistry on dust grains and in large molecules, to emission lines from both molecular and atomic transitions in the gas. Studies of galactic and extragalactic supernovae and their remnants, discussed below, illustrate the variety of analyses already under way with Spitzer data. Here, we touch very briefly on other Spitzer explorations of post-main-sequence topics. Blommaert et al. (2005) discuss ISO's extensive work on this subject (see other articles in Volume 119 of Space Science Reviews as well).

Spitzer has unique power to identify post-main-sequence stars for further study. For example, Soker & Subag (2005) predict that there should be a large population of undiscovered planetary nebulae in the plane of the Milky Way, and Cohen et al. (2005) demonstrate the power of the GLIMPSE survey to reveal them. Spitzer enables studies of post-main-sequence stars in a variety of extragalactic environments, such as the circumnuclear region in M31 (see images in Gordon et al. 2006). Characterizing post-main-sequence evolution in nearby low-metallicity galaxies such as the Magellanic Clouds will allow sampling of an earlier epoch in our own Galaxy. Examples include the study of known planetary nebulae in the Magellanic Clouds (J. Bernard-Salas, J.R. Houck, P.W. Morris, G.C. Sloan, S.R. Pottasch, D.J. Barry, in preparation), the search for LMC RCrB stars by Kraemer et al. (2005), and the IRS spectral atlas of the LMC compiled by C. Buchanan, J. Kastner, B. Forrest, S. Raghvendra, M. Egan, D. Watson, et al., (in preparation).

---

[1]Note that high spatial resolution spectra with large ground-based telescopes have found spatial variations of the dust spectrum and crystallinity with position across large disks around A type stars (Weinberger, Becklin & Zuckerman 2003; van Boekel et al. 2004).

In addition to its role in discovering and classifying post-main-sequence stars, Spitzer is revealing interesting aspects of their evolution and the chemistry in their mass-loss shells. Hora et al. (2004) examine IRAC images of planetary nebulae that effectively locate hot dust and trace large structures of molecular gas identified through molecular hydrogen emission within some of the IRAC photometric bands. Su et al. (2004) study the planetary nebula NGC 2346 in the MIPS bands, finding interesting changes in morphology with increasing wavelength. These include a circumstellar ring that becomes optically thin at 70 μm and, most significantly, a shell of cold dust from previous mass-loss episodes detected at 160 μm. Although such cold dust halos might be expected around many planetaries, so far their detection has proven elusive (W. Latter, private communication).

IRS spectra are discovering unexpected aspects of the mineralogy and chemistry in circumstellar material. For example, Markwick-Kemper, Green & Peeters (2005) report possible oxygen-rich material in the carbon-rich outflow of the Red Rectangle; conversely, Jura et al. (2006) report carbon-rich material (PAHs) orbiting the oxygen-rich red giant HD233517 and suggest that the material may have been synthesized in situ in a long-lived disk created when the star engulfed a companion. Sloan et al. (2006) report unexpected structure in the silicate emission from the LMC Mira variable HV 2310, partly attributable to crystalline grains. In addition, Spitzer observations of mid-infrared fine structure and recombination lines have been used to diagnose ionization conditions, wind properties, and abundances in objects ranging from planetary nebulae (J. Bernard-Salas, J.R. Houck, P.W. Morris, G.C. Sloan, S.R. Pottasch, D.J. Barry, in preparation) to dramatically mass-losing stars such as Sakurai's Object (Evans et al. 2006) and Wolf-Rayet stars (Morris, Crowther & Houck 2004; Crowther, Morris & Smith 2006).

## 3.4. Supernovae

Infrared observations of supernovae and supernova remnants can provide important insight into the origin and dissemination of heavy elements, both by detecting fine-structure emission lines from these materials and by enabling the study of dust created in and heated by the supernovae. The infrared spectrum of a supernova remnant also provides samples of pre-existing circumstellar and interstellar matter and can be diagnostic of the interaction of the supernova with its environment. Dwek & Arendt (1992) report that IRAS detected ~30% of the known galactic supernova remnants, providing evidence for collisional (as opposed to radiative) heating of the remnant dust and for grain processing in some remnants. All of the detected remnants radiate more power in the infrared than in the X-ray. More recently, ISO has studied the Kepler, Tycho, Cas A, and Crab Nebula remnants (Douvion, Lagage & Cesarsky 1999, Douvion et al. 2001), finding a range of properties, as expected from the known variety of supernovae. In Cas A, Arendt, Dwek & Moseley (1999) find a rich infrared emission-line spectrum and a broad 22-μm emission feature, indicative of a class of silicate minerals different from that typically associated with interstellar dust grains.

The explosion of SN1987A in the LMC provided a unique opportunity for detailed infrared observations of the earliest stages of a supernova (Wooden et al. 1993, Wooden 1995). A rich emission spectrum including contributions from Fe, Ni, Co, Ar, CO, and SiO persisted for more than 400 days, and the shift of the bolometric luminosity into the infrared following the onset of dust formation was recorded ~530 days after the explosion. Bouchet et al. (2004) detect the faint infrared glow of the warm dust in the remnant more than 16 years after the event and present mid-infrared images of the previously ejected circumstellar ring now being shock-heated by the supernova ejecta.

Many of the broad lines in the SN1987A spectrum had brightness $\gtrsim 10$ Jy even a year after the explosion. The IRS has measured sources fainter than ~1 mJy; thus Spitzer can study in detail supernovae in galaxies as distant as ~10 Mpc. Barlow et al. (2005), Kotak et al. (2005), and Stanimirovic et al. (2005) have already reported Spitzer results on extragalactic supernovae. For supernova remnants within our galaxy, Spitzer's large-scale maps have already yielded surprising results, and Spitzer can obtain detailed spatially resolved spectra of both gas phase and dust emission.

**3.4.1. IMAGING OBSERVATIONS** Within a supernova remnant, a small amount of mass condenses into small, silicate-rich grains and other refractory materials such as aluminum oxide. During the first few years after the supernova explosion, this dust may be heated by the radioactive ejecta of new heavy elements from the supernova (Bouchet et al. 2004 and references therein). Later, the emission—with contributions from both freshly synthesized and pre-existing grains—arises from very small grains stochastically heated by the energetic electrons also responsible for the thermal X-ray emission (Dwek & Arendt 1992). Hines et al. (2004) report *Spitzer* images of the young supernova remnant Cas A, which they discuss in terms of this process. They estimate that a total dust mass of only ~3 × $10^{-3}$ $M_\odot$ accounts for the mid-infrared emission from the remnant. They also call attention to a region at the edge of Cas A where a shock wave produced by the expanding supernova remnant excites infrared radiation from the dust in an adjacent molecular cloud.

A far larger mass of cold dust— about 3 $M_\odot$—had been reported from submillimeter observations of Cas A (Dunne et al. 2003). However, Krause et al. (2004) use a combination of Spitzer data and millimeter-wave molecular line observations to argue that this material is foreground interstellar gas and dust, a result confirmed by Wilson & Batria (2005). Stanimirovic et al. (2005) report observations of 1E 0102.2-7219, a ~1000-year-old supernova remnant in the Small Magellanic Cloud similar to Cas A. They estimate ~8 × $10^{-4}$ $M_\odot$ of hot dust is visible in this object and point out that this value is ~100 times below recent theoretical predictions; Spitzer observations of young remnants should explore this discrepancy futher.

Lee (2005) and Reach et al. (2006) search the GLIMPSE data for supernova remnant detections. Each group detects about 20% of the supernova remnants known in the radio band within the GLIMPSE survey region. Reach et al. (2006) show that a broad variety of emission mechanisms dominate the various IRAC bands.

**3.4.2. INFRARED ECHOES** Supernovae can also emit through infrared echoes, when nearby dust is heated by the outgoing energy flash from the explosion (e.g., Douvion et al. 2001). A Type II progenitor loses material in a ~1500-km s$^{-1}$ wind at rates of $10^{-6}$ M$_\odot$ per year. This wind will sweep up any surrounding material to generate a wind-blown bubble. Van Marle, Langer & García-Segura (2004) calculate that by the time a 25-M$_\odot$ star becomes a supernova it can have blown a bubble of ~35-pc radius into a surrounding interstellar medium of density $3 \times 10^{-23}$ g cm$^{-3}$. Thus, for up to ~100 years after the explosion, we can expect an infrared echo generated in material that the star has lost into this bubble, as in SN 2002hh in NGC 6946. Barlow et al. (2005) model the *Spitzer* observations of this event and show that they can be explained as emission by 0.1–0.15 M$_\odot$ of dust (and hence some 10 M$_\odot$ of dust and gas) ejected by the progenitor prior to the explosion.

An infrared echo can also arise as the flash propagates through the surrounding interstellar medium (ISM), including any stellar ejecta that have penetrated into the ISM, as found for Cas A by Krause et al. (2005). In this case, *Spitzer* traces the progress of the flash through the ISM, resulting in the illusion of speed-of-light motions as the expanding flash heats various cloudlets. The pattern of apparent motions close to the remnant (Figure 10) suggests a more recent event than the main explosion, possibly a giant flare from the central neutron star occurring around 1952. This hypothesis will be tested by tracking these motions throughout the *Spitzer* mission.

**3.4.3. INFRARED SPECTRA** T.L. Roellig, T. Onaka & K.-W. Chan (in preparation) present complete *Spitzer* 5.2–36-μm spectra of a bright knot in the Kepler remnant showing emission lines of Ni, Ar, S, Ne, and Fe superposed on an underlying continuum. The emission line spectrum is consistent with J-shock heating of material with interstellar abundances (see Hollenbach & McKee 1989). The continuum can be approximated by emission from collisionally heated dust grains, but a detailed fit will require more sophisticated models.

Kotak et al. (2005) measured the Ni II line at 6.62 μm in SN 2004dj in NGC 2403 at a distance of 3.1 Mpc. The measured line intensity implies a stable nickel mass of at least $2.2 \times 10^{-4}$ M$_\odot$ and suggests that the progenitor to the supernova was a red supergiant with initial mass ~10–15 M$_\odot$. Kotak et al. (2005) also review the advantages of mid-infrared fine-structure lines for determining heavy element abundances in supernovae: (*a*) the far fewer line transitions in the infrared result in less blending; (*b*) the infrared lines have greatly reduced sensitivity to extinction; and (*c*) the infrared line strengths are insensitive to temperature.

## 3.5. Brown Dwarfs

The initial detections of brown dwarfs occurred in the 1990s. Since then, systematic surveys reviewed by Kirkpatrick (2005) have shown that brown dwarfs compare in number with all other stars in the neighborhood of the Sun. Brown dwarfs are divided into three main subclasses, L, T, and Y; L-class objects (1400–2600 K) are below the hydrogen-burning limit with no methane, T-class objects (600–1400 K) are below the hydrogen-burning limit with methane, and the as-yet undiscovered Y-class objects (150–600 K) would be cool enough to have water clouds. As all brown dwarfs steadily cool during their lives, there is no equivalent to a main sequence, and an object's spectral type and mass do not correspond uniquely. Even Jupiter, with mass ~0.0015 M$_\odot$, has a substantial residual internal heat source and falls into the brown dwarf category by some determinations.

Spitzer can study brown dwarfs in detail, both photometrically and spectroscopically. Searching for and studying brown dwarfs in young clusters has proven particularly fruitful for Spitzer investigations, as discussed in Section 3.1.4. The discussion here focuses on Spitzer's studies of older, field brown dwarfs.

**3.5.1. SPECTROSCOPY OF FIELD BROWN DWARFS** IRS can obtain mid-infrared spectra of field brown dwarfs for comparison with model atmospheres of very low-mass objects. Roellig et al. (2004a) observed an M star, an L dwarf, and a T dwarf binary system, and report excellent agreement between their observed spectra and the most recent atmospheric models. Roellig et al. (2004a) also reports the first detection of ammonia in the atmosphere of a brown dwarf, as well as the first detection of the predicted methane feature located at 7.8 μm. Currently available data permit the construction of a mid-infrared MLT spectral sequence (Cushing et al. 2006). Figure 11 shows objects in this sequence ranging from early M dwarfs to cool T dwarfs. As predicted by the models, the spectra become progressively more complicated as the effective temperature falls. The region around the transition between the L and T spectral types is particularly interesting, because the models predict that atmospheric clouds of silicates and iron should become directly visible in the mid-infrared in these objects. At higher effective temperatures the clouds do not form, whereas at lower effective temperatures the clouds are physically located below the photosphere. In that location, they are not directly observable, although their influence on the atmospheric energy transport can affect the observable photosphere above. The *Spitzer* data clearly demonstrate the effects of cloud opacity in the L–T transition regime, as evidenced by a flattening in the atmospheric gas-phase absorption features, due to the much broader solid state features from the dust clouds.

These data support detailed comparison of spectra and models. Figure 12 shows an IRS spectrum of Gl570D, a T8 dwarf with an effective temperature of 800 K. The strong methane, water, and ammonia absorption bands are immediately visible in this spectrum. Two model atmospheres are overlaid on these data, one with and one without clouds. The agreement is reasonable except for the region between the ammonia bands near 10.6 μm. The similarity of the model spectra with and without clouds shows that for these very low temperature objects, any clouds are located too far below the photosphere to greatly affect the mid-infrared spectrum. More careful analysis of the spectra shows that the spectral features are indicative of nonequilibrium chemistry in the brown dwarf's atmosphere. The IRS should be capable of studying the atmospheres of even cooler Y-dwarfs when or if they are discovered.

**3.5.2. WHITE DWARF–BROWN DWARF BINARY SYSTEMS** Farihi, Zuckerman & Becklin (2005) used *Spitzer* to identify an unresolved brown dwarf companion to the white dwarf GD 1400. Because the brown dwarf dominates the infrared radiation from the system, it should be possible to obtain spectra of such companions to compare with those of the field brown dwarfs and perhaps to illuminate the evolution of the parent binary system.

## 4. PLANETARY SYSTEM FORMATION AND EVOLUTION

During and long after the dissipation of a protoplanetary disk, collisions of asteroids or evaporation of comets around a main sequence star inject dust into its circumstellar environment. We refer to the tenuous disk that forms and regenerates through this process as a planetary debris disk. With rare exceptions, these systems are too diffuse and faint for detection, let alone detailed study, by warm telescopes. The diagnostic capabilities of IRS augment the photometric capabilities of IRAC and MIPS to take us far beyond the simple identification and characterization of such systems, enabling detailed studies of disk composition and structure.

The infrared is the prime wavelength band for direct detection of extrasolar planets because the planet/star flux ratio, though small, should be considerably larger than in the visible spectrum. The sensitivity and stability of Spitzer's instruments have allowed full exploitation of this advantage, leading to the direct detection of radiation from several extrasolar planets with orbital geometries particularly favorable for this purpose.

Spitzer studies of both extrasolar planets and debris disks can be correlated directly to phenomena in our own solar system. The extrasolar planets appear similar in mass and size to Jupiter; the debris disks are exactly analogous to solar system dust structures such as the zodiacal cloud; and the objects that feed the debris disks have local counterparts in the comets, asteroids, KBOs, and other small bodies within the solar system. Spitzer has begun systematic studies of these solar system phenomena.

### 4.1. Evolution of Debris Disks

Debris disks are produced from larger bodies and respond gravitationally to the planets whose space they share, so their properties can reveal the character of extrasolar planetary systems. Overall, the dust density and rate of dust production decline with time, as radiation pressure or radiation drag remove the small particles from the system and the parent planetesimals are no longer replenished. In the current-day solar system, the level of debris generation is very low. The asteroid belt has a fractional luminosity $f \leq 10^{-7}$. It was therefore a surprise when the IRAS mission demonstrated that a number of nearby stars have excess emission with $f \geq 10^{-4}$ (Aumann et al. 1984). A second surprise was that these systems were in orbits with radii of roughly 100 AU. This discovery, made prior to our finding the first KBOs, exemplifies how debris disks can reveal aspects of other planetary systems relevant to understanding the solar system.

Less than $10^{-2}$ Earth masses of dust are required to produce $f \sim 10^{-4}$. Although the dust is much colder than the star, the surface area of the finely divided small particles is large enough that their far-infrared radiation dominates that of the star by factors of 50 to 100 at 60 μm and 100 μm for the prominent disks discovered by IRAS (see Backman & Paresce (1993) for a discussion of these and other debris disk fundamentals).

**4.1.1. SPITZER'S ROLE** IRAS and ISO show that debris disks with $f \gtrsim 10^{-5}$ occur around perhaps 15% of nearby main sequence stars of spectral types K through A. However, the formal uncertainties in this estimate are large. These missions also established a rough timescale of a few hundred million years for the decay of the systems around young stars (Habing et al. 2001, Spangler et al. 2001, Decin et al. 2003), but found that some much older stars also have large excesses (Decin et al. 2003). A general summary of debris disk studies after the completion of these two missions can be found in Zuckerman (2001) and Caroff et al. (2004). HST has imaged scattered light from a number of circumstellar disks, but with the exception of the one around Fomalhaut (Kalas, Graham & Clampin 2005), successful imaging has been achieved only for very young systems. In the submillimeter spectral range, a few of the nearest debris disks have been imaged (Wyatt et al. 2003 and references therein) and a few others have been measured photometrically. In most cases these imaging experiments have established that the dust in these systems has a disk- or ring-like distribution, as is assumed throughout this discussion.

*Spitzer*'s contributions derive largely from its ability to search a virtually unlimited number of stars for debris-derived infrared excess emission. At 24 μm, *Spitzer* can detect as little as 10% above the bare photosphere, limited by our ability to estimate the intrinsic photospheric emission. At 70 μm, *Spitzer* can measure levels 20–30% above the photosphere. Table 2 summarizes many of the debris disk surveys carried out with *Spitzer* to date. With spectroscopy, such searches can be expanded both to constrain the location of the debris and to probe its mineralogical content. *Spitzer*'s relatively high angular resolution allows imaging of prominent, nearby debris systems as well as improved discrimination against spurious debris disks caused by stellar heating of nearby interstellar material or a background source in the large IRAS beam (e.g., Kalas et al. 2002).

**Table 2**  Published Results from Spitzer Debris Disks Surveys Carried Out Mainly at 24 and 70 μm

| Reference[a] | Sample | 24-μm-only excesses | 70-μm-only excesses | 24- & 70-μm excesses | Range of $f$ | Comments |
|---|---|---|---|---|---|---|
| Beichman et al. 2005 | 26 FGK stars with RV companions, older than ~1 Gyr | none | 6 (0.23) | none | $>1.2 \times 10^{-4}$ to $>1 \times 10^{-5}$ | |
| Bryden et al. 2006 | 69 FGK stars with median age ~4 Gyr | 1 (0.01) | 7 (0.10) | none | $>6 \times 10^{-6}$ to $>6.8 \times 10^{-5}$ | |
| Chen et al. 2005a | 40 F and G stars in Sco-Cen OB group. Ages 5-to-20 Myr | 7 (0.17) | none | 7 (0.17) | $4 \times 10^{-5}$ to $3.00 \times 10^{-3}$ | no photospheres detected at 70 μm |
| Chen et al. 2005b | 39 A through M dwarfs. Ages 12-to-600 Myr | 1 (0.026) | 4 (0.10) | 2 (0.05) | $2 \times 10^{-6}$ to $7.70 \times 10^{-4}$ | |
| T.N. Gautier III, G.H. Rieke, J. Stansberry, G.C> Bryden, K.R. Stapelfeldt, et al. (in preparation) | 30 local M stars not thought to be young | none | none | none | NA | see text |
| Gorlova et al. 2004 | 63 stars in M47, age ~100 Myr, types A through M | 11 (0.17) | NA | NA | NA | mix of objects includes several late-type debris disk candidates |
| Kim et al. 2005 | ~35 solar type stars from the FEPS sample | 1 (0.03) | 5 (0.14) | none | $2.3 \times 10^{-4}$ to $<3 \times 10^{-5}$ | 70-μm only detections aged ~0.7–3 Gyr. IRS data used as well |
| Low et al. 2005 | 24 stars in TW Hydra association, age ~8–10 Myr, types A through M | none | 1 (0.04) | 5 (0.21) | 0.27 to $8.60 \times 10^{-4}$ | includes objects still in accretion phase. Some 160 μm detections reported as well |
| Rieke et al. 2005 | 266 A stars aged <10-to-800 Myr | 72 (0.27) | 70-μm data on this sample in preparation by K.Y.L. Su et al (see text) | | | Max 24 μm excess declines as 150 Myr·t$^{-1}$ |
| Stauffer et al. 2005 | 20 G dwarfs in the Pleiades, age ~100 Myr | 3 (0.15) | none | none | $1.6 \times 10^{-4}$ or less | no photospheres detected at 70 μm. IRS data used as well |

[a]Use this table to point to papers and programs of particular interest to you. In general, the results of each study are too complex to be summarized completely in tabular form. Studies summarized here are primarily based on MIPS photometry at 24 and 70 μm. Beichman et al. (2006) show that Low Resolution IRS Spectra can pick up disks at levels of $f$ or in wavelength regions difficult to achieve with photometry alone and use the technique to set limits on asteroidal emission from stars with radial velocity companions.

**4.1.2. SPECTRAL PROPERTIES** The broad SEDs of the great majority of debris disks are remarkably similar, with shapes indicating that the material is at a temperature of about 70 K and therefore 10–100 AU from the star. Hence, these regions appear to be analogous to the Kuiper Belt. For old stars of roughly solar type (late F to early K), *Spitzer* spectra generally show no detectable debris emission at wavelengths short of 33 μm (Kim et al. 2005, Beichman et al. 2006), indicating that these rings of debris usually terminate at their inner edges with little material within the zone where the terrestrial planets lie in the solar system.

A-type stars behave differently; a detectable excess at 24 μm almost always accompanies a large excess at 70 μm (K.Y.L. Su, et al., in preparation). However, the two well-imaged A-star debris systems, Fomalhaut and Vega, show Keplerian systems of debris at radius ~100 AU in the submillimeter range (and also in the far-infrared for Fomalhaut). Thus, the 24-μm emission arises from tenuous material inside the Kuiper-Belt analog region. That is, despite the differences in SED, the A stars may not host qualitatively different structures from the debris around cooler stars.

Only a small number of extreme systems have been found that depart substantially from this overall pattern. For example, *Spitzer* has identified only a few cases of 24-μm-only excesses, which might suggest asteroid dust, as appears to be the case for Zeta Lep (cf. Chen & Jura 2001). A particularly interesting example, the ~2-Gyr-old K star HD69830, has a 24-μm excess due almost entirely to a large population of small crystalline grains (Beichman et al. 2005a) within a few astronomical units of the star, but without a substantial reservoir of cooler material. The grains are known to be less than a few micrometers in size because of the strong emission features due to crystalline silicates in the 10-μm region (Figure 9). Models of HD69830 demonstrate that these small grains have very short lifetimes before their destruction or expulsion from the system. Their presence in such numbers indicates their recent generation by a transient phenomenon such as a supercomet or a collision in a densely populated asteroid belt (Beichman et al. 2005a). Similar conclusions may apply to other systems of this type (Chen et al. 2005a,b; Song et al. 2005; Hines et al. 2006); even among this group HD69830 may be unique with its >2 Gyr stellar age.

Uzpen et al. (2005) report a number of debris disks in the GLIMPSE database on the basis of excess emission in the 8-μm IRAC band. However, this sample may be contaminated owing to the distance of the GLIMPSE sources and the strong interstellar PAH emission at 8 μm. IRS spectroscopy can test the true nature of the excesses.

**4.1.3. DIRECT IMAGING** *Spitzer* has imaged well four debris disks as of this writing: Fomalhaut, Vega, eps Eridani, and beta Pic (however, the last is too young to be considered a mature debris system). The imaging tests the assumptions made in analyzing the SEDs of the many spatially unresolved systems explored by *Spitzer*; thus, the marked variety of behavior even within this small sample challenges any interpretation critically dependent on a particular assumption about the disk geometry. Fomalhaut does indeed behave as expected, with a circumstellar ring of radius ~110 AU prominent at 70 μm (Figure 13) and through the submillimeter, but filled in at 24 μm by particles being dragged into the star. Asymmetries in the visible image may reflect the gravitational influence of a planet orbiting interior to the ring (Kalas, Graham & Clampin 2005); at 70 μm the ring shows asymmetry consistent with that seen in the visible. Beta Pic is large in extent (K.Y.L. Su, personal communication), as indicated also by previous optical and mid-infrared ground-based imaging, consistent with arguments that grains are being ejected by photon pressure. Telesco et al. (2005) suggest from ground-based observations that a recent catastrophic planetesimal collision generated a prominent cloud of tiny grains within this disk. Given the youth of the star (10–20 Myr; Lanz, Heap & Hubeny 1995), this behavior may provide important insight into the embryo planet phase in the evolution of terrestrial planets.

*Spitzer* images, combined with those at other wavelengths, highlight the importance of processes that segregate particles by size, producing images that vary in appearance with wavelength. Eps Eridani shows a ~60-AU radius ring in the submillimeter range, but the ring is already filled at 70 μm, and the 70-μm emission extends exterior to the submillimeter ring as well (D. Backman, et al., in preparation). The Vega images are yet more striking (Figure 13). Debris can be traced to a radius of nearly 1000 AU. Su et al. (2005) fit the characteristics of the system with small grains (~10-μm radius) that are being ejected by photon pressure. The fits indicate that the grains originate in a Keplerian ring of objects detected in the submillimeter range at a radius of about 90 AU, where a large collision may have taken place on the order of a million years ago, setting up the collisional cascade responsible for the small grains. As with HD69830, the high loss rate for these grains makes it implausible that the Vega system has always had its current appearance.

**4.1.4. EVOLUTION** The new *Spitzer* results emphasize the role of individual planetesimal collisions in producing large populations of grains that can dominate the radiometric properties of these systems for a few million years. These events complicate comparisons with most of the existing models, which assume a smooth and continuous evolution (the models of Kenyon & Bromley 2004, 2005 and Grogan, Dermott & Durda 2001 are exceptions). They also make it difficult to determine to what extent the variety in observed disk properties reflects fundamental differences in the disks.

*Spitzer* confirms previous indications that strong excesses, suggesting active terrestrial planet building (and destruction), occur commonly around stars less than 100 Myr in age (Gorlova et al. 2004; Chen et al. 2005a,b; Rieke et al. 2005). A-type stars are attractive targets for tracking disk evolution because a significant 70-μm excess is almost always accompanied by a detectable excess at 24 μm (K.Y.L. Su et al., in preparation). Because of the high angular resolution of the MIPS 24-μm band, its reduced susceptibility to cirrus confusion emission, and its high sensitivity to stellar photospheres, A-star excesses ~10% above the photospheric output can be detected at 24 μm to a distance of ~500 pc, putting the nearest young stellar clusters within range. Rieke et al. (2005) report a preliminary study based on this capability, presenting a compilation of 266 A stars, some of which were observed by IRAS and ISO. In total, 27% of the stars have 24-μm excesses. Even in the youngest age range (<25 Myr), ~50% of the stars show no excesses, while some young stars have extremely large excesses. As shown in Figure 14, the envelope of the maximum excess decays with age, $t$, roughly as $150 \text{ Myr} \cdot t^{-1}$.

Again, these results highlight the stochastic evolution of debris disks, as stars more than 1 Gyr in age would hardly ever have significant excesses according to the 150 Myr·$t^{-1}$ enveloping behavior, yet many of them do. In the combined results from Kim et al. (2005) and Bryden et al. (2006), 13±4% of old, solar-like (late F to early K) stars in the solar neighborhood have substantial excess emission at 70 μm, although virtually none have excesses at 24 μm. The limiting $f$ for the F-to-K stars is ~$10^{-5}$.

In evaluating these debris-disk differences among stellar types, we need to distinguish between effects due to age and those due to stellar mass. A typical solar-type debris system circles a star with a main sequence lifetime exceeding that of an A star. Observations of solar-type stars in the Pleiades, at 120 Myr, indicate a significant incidence (~15%) of 24-μm excesses (Stauffer et al. 2005; N. Gorlova, private communication). It appears to take 100–200 Myr for debris to clear from the terrestrial planet zone in a typical planetary system, independent of stellar mass between solar and A-type. Jura (1990), using IRAS data, showed that excesses are rare in G giants, former A stars that have left the main sequence. This suggests that the high rate of A-star detection arises in part because of their youth, and the slow clearing of debris systems in general.

**4.1.5. DEPENDENCE ON OTHER FACTORS** Detection of debris to a given level of $f$ becomes more difficult as the stellar luminosity and temperature decrease along the main sequence. The stellar luminosity is directly proportional to $T^4$ (where $T$ is its temperature), whereas the Rayleigh-Jeans photospheric flux density above which the infrared excess must be detected is directly proportional to $T$. Thus, for a given contrast limit above the photosphere, the limiting $f$ will scale as $T^{-3}$.

Debris disk excesses in M stars have been elusive, because of this increase in detection difficulty with decreasing stellar luminosity and because of the intrinsic faintness of these targets. The only M stars known to have excesses at 25 μm and beyond are T Tauri stars and young objects such as AU Mic (Song et al. 2002; Kalas, Liu & Matthew 2004). T.N. Gautier III, G.H. Rieke, J. Stansberry, G.C. Bryden, K.R. Stapelfeldt, et al. (in preparation) have studied 30 nearby early M stars not expected to be particularly young and find no definite excesses at either 24 or 70 μm. They also show that the average excess over the photosphere for these stars at 70 μm is at least a factor of ~6 lower than the average excess level for the F-, G-, and K-star sample of Bryden et al. (2006); by the argument given above, however, the M stars may still have $f$ comparable to that of their more massive counterparts.

Bryden et al. (2006) show the incidence of debris disks to be roughly independent of stellar metallicity, in dramatic contrast to the rapid increase in radial velocity planet detections with increasing metallicity (Fischer & Valenti 2005). Beichman et al. (2005b) report 70-μm excesses around 6 of 26 FGK stars with known radial velocity planetary companions, a rate of 23%. Supplementing this result with additional ongoing GTO and GO observations (G. Bryden, personal communication) yields a net of 7 excesses in 37 stars with known planets, or a rate of 19±7%. Considering the small sample, this value does not differ significantly from the detection rate of 13±4% for similar, single stars (Kim et al. 2005, Bryden et al. 2006). J.A. Stansberry & D.E. Trilling (personal communication) also find a similar detection rate for 70-μm excesses around binary stars. The incidence of detectable excesses thus appears surprisingly independent of metallicity or the presence of companions.

*Spitzer* has identified debris-disk-like structures in a surprising variety of environments. The white dwarf Giclas 29-38 shows an infrared excess most naturally attributed to a circumstellar dust cloud or debris disk (Chary, Zuckerman & Becklin 1999). Reach et al. (2006) have obtained *Spitzer* observations of this system—including the first infrared spectra—that strongly support this identification. The spectrum shows a strong emission feature at 10 μm, well modeled as emission from small carbonaceous and silicate particles, which would be located just a few solar radii from the star. A possible explanation is that comets and/or asteroids survived the post-main-sequence evolution of the main sequence progenitor (estimated to have been a ~3-$M_\odot$ A star) and have now been disrupted to produce the cloud of particles radiating at 10 μm. At the other end of the mass spectrum, Kastner et al. (2006) identify debris-disk analogs around two highly luminous B[e] stars in the LMC. They suggest that the grains may arise in Kuiper Belts around these stars while pointing out that such structures might have difficulty surviving the luminous main sequence lifetime of the O-star progenitors of the B[e] stars. Finally, Wang, Chakrabarty & Kaplan (2006) report *Spitzer* detection of excess infrared emission from an X-ray pulsar identified as a neutron star. They attribute the infrared radiation to a "fallback disk" produced around the neutron star by ejecta from the original supernova explosion.

**4.1.6. COMPARISON WITH THE SOLAR SYSTEM** From the crater record on the Moon and other arguments, we know that Earth grew up in a dangerous neighborhood. The heavy bombardment of the Earth and Moon ended about 700 Myr after the formation of the Sun (see Gomes et al. 2005 and references therein). Throughout the period of bombardment, there would have been a continuous production of debris, probably with spikes in grain production whenever a particularly large collision occurred. Strom et al. (2005) attribute the period of heavy bombardment to dynamical instabilities triggered by the orbital migration of the giant planets. This raises the intriguing possibility that the sporadic outbursts producing the stochastic behavior of the planetary debris disks described above may trace the dynamical evolution of their host planetary systems.

The Kuiper Belt is the region within the Solar System most analogous to the debris disks around nearby stars. As summarized in Kim et al. (2005), Backman, Dasgupta & Stencel (1995) used COBE data to place an upper limit of $f = 10^{-6}$ on the debris system associated with the Kuiper Belt. Models by Moro-Martín & Malhotra (2003) suggest that the actual level of emission should only be a factor of a few below this limit. Thus, the debris disk emission of the solar system probably falls about two orders of magnitude fainter than the faintest systems being detected with *Spitzer*. However, the observations are consistent with the detected systems defining the high end of a statistical distribution of debris disks including, near the average level, the one in the solar system (Kim et al. 2005, Bryden et al. 2006). That is, the detected systems appear to have a strong family resemblance to the debris in our own system.

*Spitzer*'s studies therefore have direct relevance to understanding the conditions under which potentially habitable planets form and evolve around hundreds of nearby stars. The events and processes occurring in these systems must have analogs to those that shaped the solar system. This connection was fragmentary and incomplete prior to the launch of *Spitzer*, but it is now firmly established.

## 4.2. External Planet Occultation Results

The first confirmed extrasolar planet orbiting a solar-type star—in the 51 Pegasi system (Mayor & Queloz 1995)—was also the first example of a "hot Jupiter" or "roaster," that is, a planet with a mass comparable to that of Jupiter that orbits within ~0.05 AU of its parent star. The origins of these planets and their orbits remain a puzzle; however, their temperatures (close to 1000 K) and the amount of infrared radiation they produce should make them detectable by *Spitzer* out to distances of 200 pc.

To do this, Charbonneau et al. (2005) and Deming et al. (2005) exploited the favorable case where the orbital inclination permits mutual eclipses of the star and the planet. The disappearance of the planet behind the star during the secondary eclipse leads to a small drop in the infrared radiation from the star-planet system. As shown in Figure 15, Deming et al. (2005) used MIPS to detect the secondary eclipse in the HD209458 system with an amplitude at 24 μm of 0.26±0.046%, whereas Charbonneau et al. (2005) used IRAC to detect the eclipse in the TrES-1 system with amplitude 0.066±0.013% at 4.6 μm and 0.225±0.036% at 8 μm. Such measurements provide unique constraints on the temperature, Bond albedo, and perhaps the composition of the planets, which in turn provide clues to how they formed. In addition, the *Spitzer* data determine the time of the secondary eclipse precisely, which constrains the eccentricity of the orbit. In the two cases studied to date, the temperature of the planets was found to be ~1100 K; for TrES-1, the Bond albedo was found to be 0.31±0.14. In each case, the timing of the secondary eclipse suggests that the orbital eccentricity of the planet is too low for tidal dissipation to be a significant heat source, ruling out tidal heating as an explanation for the unexpectedly large radius of the planet orbiting HD209458.

We can anticipate further results of this type from *Spitzer,* including studies of additional transiting planets. However, the measurements of TrES-1 and HD209458 have already led to interesting speculations about these planets. As one example, Burrows, Hubeny & Sudarsky (2005) show that agreement with the *Spitzer* results appears better for a model in which the incident starlight is reradiated primarily by the day side—as seen in the eclipse measurements—rather than uniformly by the whole disk. On the other hand, Barman, Hauschildt & Allard (2005) report that significant day-to-night energy redistribution is required to reproduce the observations. Resolution of this issue, which may be possible with additional *Spitzer* data, will clarify the nature of the mass motions in the extrasolar planets' atmospheres that would be required to redistribute the incident energy. Looking at the *Spitzer* data from a different perspective, Fortney et al. (2005) suggest that the very red 4.5–8 μm color of TrES-1 results from a metallicity enhancement by a factor of 3–5 in the planet over the parent star; this may speak to the mode of formation of the planet. Other issues that *Spitzer* observations will illuminate include the chemical and molecular composition of the planets' atmospheres and, perhaps, the presence or absence of clouds or dust within them.

### 4.3. Solar System Studies

*Spitzer*'s capabilities for the study of small bodies in the solar system—asteroids, comets, planetary satellites, and KBOs—take on new significance with the flood of data being returned on extrasolar planets and planetary systems. Combining *Spitzer* studies of the solar system with observations of extrasolar planetary systems opens the exciting possibility of comparative planetology studies, in which we can apply our understanding of our own solar system to analysis of the extrasolar systems. At the same time, the full range of extrasolar systems accessible to *Spitzer*, together with the favorable external and complete views we have of them, may improve our understanding of our home planetary system.

**4.3.1. ASTEROIDS** The ecliptic plane survey portion of the First Look Survey (Meadows et al. 2004) has made measurements over 0.13 square degree areas at 8 and 24 μm at a solar elongation of 115° and ecliptic latitudes of 0° and 5° that are sensitive to main belt objects ~0.5 km in diameter. In the ecliptic plane, the results translate to an asteroid number density of 154±37 per square degree brighter than 0.1 mJy (5-σ) at 8 μm. This number is consistent with the low side of the predictions of several previous asteroid population models. Interestingly, the *Spitzer* counts appear to fall less steeply with ecliptic latitude than predicted by the models, suggesting that the subkilometer asteroids studied for the first time by *Spitzer* at these wavelengths may be more broadly distributed in ecliptic latitude than are the larger objects.

The Trojan asteroids, gravitationally trapped in Jupiter's stable Lagrange points, may be among the most primitive objects in the solar system. J.P. Emery, D.P. Cruikshank & J. Van Cleve (submitted) have used the IRS low-resolution modules from 5.2 to 37 μm to obtain the first thermal infrared spectra of Trojan asteroids. The spectra show broad spectral features suggesting that fine-grained silicates—both crystalline and amorphous—cover the surfaces of these objects. These results support the suggestion that the red spectral reflectivity of these objects is due to silicates rather than to organic material.

The spectrum of the asteroidal emission—in excess of the predictions of the standard thermal model—is strikingly similar to that of comet Hale-Bopp and of the debris disk orbiting the nearby solar-type star HD69830 (Figure 9). *Spitzer* observations of comet Schwassmann-Wachmann 1 (SW-1), which has a semimajor axis of ~6 AU, show similar features (Stansberry et al. 2004). Together, the results suggest that the middle part of the solar system, the transition region between rocky and icy objects, may not contain an abundance of organic materials.

**4.3.2. COMETS** *Spitzer*'s ability to measure thermal emission from comets at large heliocentric distances permits determination of the temperatures, sizes, and albedos of cometary nuclei. Stansberry et al. (2004) found SW-1 to have a nuclear radius of 27 km and an albedo of 0.025. Lisse et al. (2005) studied comet Tempel 1 when it was about 3.7 AU from the Sun. These measurements determined that the nucleus has semimajor and semiminor axes of ~7.2±0.9 km and 2.3±0.3 km, respectively, and that the albedo is 0.04±0.01. The very low albedo for these cometary nuclei, which is consistent with other determinations (Lamy et al. 2004), contrasts sharply with the high albedos determined by *Spitzer* for a number of large KBOs and trans-Neptunian objects (see Section 4.3.3).

*Spitzer* joined many ground- and space-based telescopes in observing the Deep Impact encounter with comet Tempel 1 on July 4, 2005. Before impact, a smooth ambient coma surrounded the nucleus; afterwards, ejecta from the comet flowed out at velocities ~200 ms$^{-1}$ for 40–50 hours. Lisse et al. (2006) use IRS spectra to identify an intriguing variety of materials within the ejecta from the impact, including clay, crystalline silicates, and carbonates as well as polycyclic aromatic hydrocarbons (see Figure 16). The structure in the spectrum of the ejecta contrasts sharply with the smooth spectrum of the cometary nucleus prior to the encounter (Lisse et al. 2005) and is similar to, but with even stronger features than, the spectra of comet Hale-Bopp and the debris disk around HD 69830 (Figure 9). The *Spitzer* results show that Tempel 1 contains materials characteristic of a wide range of temperatures and water content, suggesting that the comet somehow managed to agglomerate from material that formed throughout the solar system.

Cometary mass loss within our own solar system connects with the studies of planetary debris disks, providing a local and dramatic example of the processes that maintain the extrasolar systems. *Spitzer* can connect these processes by studying not only the relatively bright comae and dust tails of active comets but also the debris trails found by IRAS to fill cometary orbits as a result of the accumulation of dust lost over many orbital periods (Reach et al. 2005; Gehrz et al. 2006). These extended, low-surface-brightness trails are not readily seen with warm telescopes.

**4.3.3. KUIPER BELT AND TRANS-NEPTUNIAN OBJECTS** Intensive studies during the past decade (see Luu & Jewitt 2002) have identified hundreds of KBOs in the region beyond Neptune, which was previously known to be inhabited only by Pluto and its satellite, Charon. These objects have been discovered through visible surveys, which provide orbital parameters and have identified dynamical families. However, the visible observations provide limited information about the physical properties of individual KBOs, because of the degeneracy between the albedo and the size of an object seen only in reflected light. *Spitzer*'s ability to measure thermal emission from KBOs breaks this degeneracy and allows determination of the albedos and radii of individual KBOs. For the KBO 2002 AW197, Cruikshank et al. (2005) use MIPS photometry to determine an albedo of 0.17±0.03 and a diameter of 700±50 km, larger than all but one main belt asteroid. They also suggest that the surface of 2002 AW197 has low but nonzero thermal inertia, indicating either a very porous surface or the presence of a material with low intrinsic thermal inertia, such as amorphous $H_2O$ ice. Additional information is available in the not infrequent case of a binary KBO. For the binary KBO 1999 TC36, Stansberry et al. (2006) determine a system albedo of 0.08 and an effective diameter of 405 km from MIPS photometry. The average density for the two components, using the system mass determined from HST observations of the binary orbit and considering the uncertainties in the albedo and the diameter, lies between 0.3 and 0.8 g cm$^{-3}$. A higher and perhaps more plausible density would be indicated if the primary were itself a binary.

M. Brown and colleagues recently announced the discovery of three unusually large and bright KBOs (Brown, Trujillo & Rabinowitz 2005; also M.E. Brown, et al. in preparation). All three have been studied with *Spitzer*. *Spitzer* detected the two smaller objects, 2003 EL61 and 2005 FY9; they appear to have albedos ~0.7 and radii ~900 km (about 75% the size of Pluto). The larger object, 2003 UB313, is more distant and was not detected by *Spitzer*. If its albedo is comparable with that of the other two objects, which would be consistent with *Spitzer*'s upper limit on its flux, it is about 20% larger than Pluto and would qualify for consideration as the tenth planet. All three of these large objects occupy high-inclination orbits and belong to the scattered Kuiper Belt population.

## 5. CONCLUSIONS

The results from *Spitzer* presented above permit the following conclusions to be stated with confidence:

1. Protoplanetary disks around solar-type stars evolve quickly, suggesting that terrestrial planet formation is well under way within less than 10 Myr. The disks dissipate from the inside out.
2. Substellar objects with mass as low as ~10 $M_{Jup}$ form by the same gravitational collapse process that gives rise to more massive stars. Their protoplanetary disks appear similar in structure and composition to those orbiting young solar-mass stars.
3. Although supernova remnants harbor only small masses of dust, large signals are detected from infrared echoes as the pulse from the supernova expansion heats surrounding circumstellar and interstellar matter.
4. Current models provide reasonably accurate predictions of the thermal infrared spectra of old, cool brown dwarfs.
5. Planetary debris disks evolve stochastically. Statistically, these systems show a gradual decline in dust content with time that can be punctuated even at late stages by events that inject large amounts of new material into the circumstellar environment.
6. The character and frequency of planetary debris disks do not vary dramatically with stellar type, metallicity, or the presence or absence of planets.
7. Kuiper Belt objects can have a wide range of albedos, from the very low values suspected prior to *Spitzer* to ~0.1–0.2 for moderate sized ones and ~0.7 for very large ones. *Spitzer* results indicate that KBOs may also have very low densities ($<1$ g cm$^{-3}$).

Looking ahead to a retrospective *Annual Reviews* chapter that might appear several years after the end of the *Spitzer* mission, we can anticipate that at least the following advances would be reported:

*Spitzer*'s ability to survey large areas and to study numerous objects in great detail—spectroscopically and photometrically—will lead to great advances in our understanding of star and planetary system formation. *Spitzer* will pin down the timescales for disk evolution, characterize the relationship between disk structure and dust composition, and clarify the importance of starless cores and infrared dark clouds in low-mass and high-mass star formation, respectively. On larger spatial scales, *Spitzer* surveys should lead to the identification of the mechanisms by which star formation propagates through large molecular complexes and permit study of the role of radiation, winds, and molecular outflows in triggering or inhibiting star formation.

*Spitzer* studies of regions of star formation may identify substellar objects with masses as low as ~1 $M_{Jup}$ forming in isolation by gravitational collapse. At the same time, *Spitzer* spectra of field brown dwarfs should permit detailed searches for small deviations from the models that may signal processes such as cloud formation and atmospheric circulation.

*Spitzer* will have studied at least several handfuls of transiting planets at a level adequate to permit detailed tests of theoretical predictions and comparisons with the giant planets in the solar system. These studies should include not only photometric measurements but also the first thermal infrared spectra of extrasolar planets. In addition, *Spitzer* spectroscopy of material in protoplanetary and planetary debris disks, as well as of comets and asteroids in the solar system, will continue to permit comparisons between the composition of extrasolar and local planetary materials. These and other examples, including *Spitzer*'s continuing studies of KBOs, will assure continued traffic across the interface between planetary science and extrasolar planet studies, to the great benefit of both disciplines.

Finally, detailed *Spitzer* studies of many phenomena, including star formation, post-main-sequence stellar evolution, and supernovae, will extend far beyond the boundaries of our own Galaxy and well in to the local universe. Just as our studies of extrasolar planetary systems illuminate phenomena within our own solar system, exploring these familiar objects in the varying environments of nearby galaxies should provide important insights into the fundamental nature of the underlying astrophysical processes.

**ACKNOWLEDGMENTS**

The authors acknowledge the contributions of many people across the project who worked with us to bring *Spitzer* to fruition. We thank L. Allen, M.J. Barlow, M. Brown, D. Charbonneau, C. Lada, K. Luhman, R. Kirshner, T. Megeath, T. Soifer, K. Stapelfeldt, J. Stauffer and, particularly, E. van Dishoeck for helpful discussions and comments on the manuscript and many colleagues for making their *Spitzer* results available prior to publication. G. Fazio acknowledges NASA support through contracts 1062296 and 1256790 issued by JPL/Caltech. G. Rieke acknowledges NASA support through contract 1255094 issued by JPL/Caltech. T. Roellig acknowledges NASA support through WBS# 420579.04.03.02.02 D. Watson acknowledges NASA support through contract 960803 with Cornell University issued by JPL/Caltech, and through Cornell subcontract 31219-5714 to the University of Rochester. M. Werner thanks Charles Alcock and the Center for Astrophysics, and Rosemary and Robert Putnam, for hospitality during the preparation of the paper. He also thanks Mary Young, Jim Jackson and, in particular, Courtney Young for editorial assistance. This research was carried out in part at the Jet Propulsion Laboratory, California Institute of Technology, under a contract with the National Aeronautics and Space Administration.

**ACRONYMS LIST**

| Acronym | Definition |
| --- | --- |
| *f* | fractional luminosity |
| IRAC | Infrared Array Camera |
| IRAS | Infrared Astronomy Satellite |
| IRS | Infrared Spectrograph |
| ISO | Infrared Space Observatory |
| KBO | Kuiper Belt object |
| MIPS | Multiband Imaging Photometer for *Spitzer* |
| PAH | polycyclic aromatic hydrocarbon |

| SED | spectral energy distribution |
|-----|------------------------------|
| SSC | *Spitzer* Science Center |
| YSO | young stellar object |

## ANNOTATED REFERENCES

Armus L, ed. 2006. *ASP Conf. Ser.* In press. The Spitzer Space Telescope: New Views of the Cosmos: proceedings of the first *Spitzer* science conference, held in Fall 2004. Includes numerous papers and abstracts reporting early *Spitzer* results

Cesarsky CJ, Salama A, eds. 2005. *Space Sci. Rev.* 119(1–4). Topical reviews of *Infrared Space Observatory* results in key science areas.

Chary R, Sheth K, Teplitz H. 2006. *ASP Conf. Ser.* In press. Infrared Diagnostics of Galaxy Evolution: proceedings of the second *Spitzer* science conference, held in Fall 2005.

Dwek E, Arendt RG. 1992. *Annu. Rev. Astron. Astrophys.* 30:11–50. Basic physics relevant to dust in supernova remnants.

Hartmann L. 2005. *ASP Conf. Ser.* 341:131. Reviews the status of disk evolution prior to the deluge of *Spitzer* data.

Kirkpatrick JD. 2005. *Annu. Rev. Astron. Astrophys.* 43(1):195–245. Up-to-date review on brown dwarfs by a leading researcher in the field.

Lada CJ. 2005. *Prog. Theor. Phys. Suppl.* 158:1–23. Summary of star-formation observations and theory pre-*Spitzer*.

Luu JX, Jewitt DC. 2002. *Annu. Rev. Astron. Astrophys.* 40:63–101. Summary of the rapidly evolving field of Kuiper Belt objects.

Reipurth B, Jewitt D, Keil K, ed. 2006. *Protostars and Planets V*. Tucson: University of Arizona: In press. Proceedings of a major conference touching on many of *Spitzer*'s science themes.

Rieke G. 2006. *The Last of the Great Observatories: Spitzer and the Era of Faster, Better, Cheaper at NASA.* Tucson: University of Arizona: In press. Comprehensive history of the *Spitzer Space Telescope.*

van Dishoeck EF. 2004. *Annu. Rev. Astron. Astrophys.* 42(1):119–67. Summarizes outstanding work from ISO on spectroscopy of interstellar and circumstellar matter. Includes wavelength regions not covered by *Spitzer* as well as higher spectral resolution observations.

Werner M. 2006. In "The Spitzer Space Telescope: New Views of the Cosmos." *ASP Conf. Ser.* In press. A Short and Personal History of the Spitzer Space Telescope.

Zuckerman B, Song I. 2004. *Annu. Rev. Astron. Astrophys.* 42(1):685–721. Discusses nearby groups of young stars that are promising candidates for study with *Spitzer.*

Zuckerman B. 2001. *Ann. Rev. Astron. Astroph.* 39:549–80. Pre-*Spitzer* review of debris disks post-ISO and –IRAS.

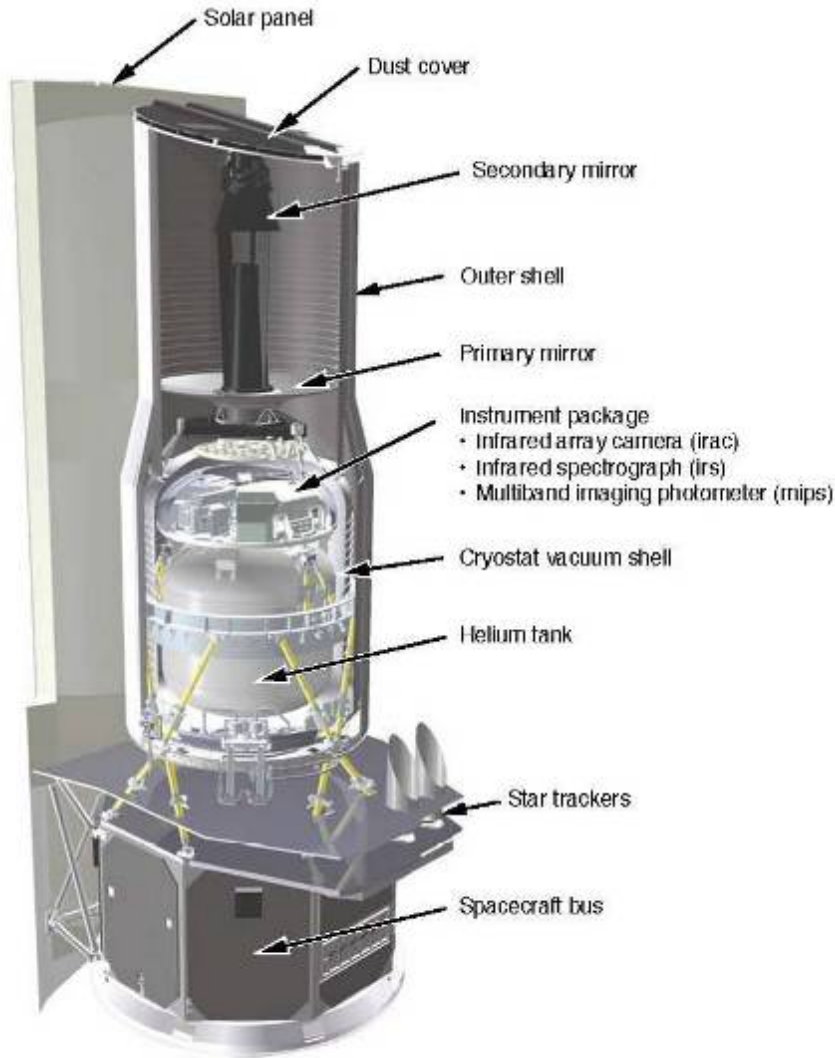

**Figure 1:** The *Spitzer Space Telescope.* The outer shell forms the boundary of the Cryogenic Telescope Assembly (CTA), which incorporates the telescope, the cryostat, the helium tank, and the three instruments, the Multiband Imaging Photometer, the Infrared Spectrograph, and the Infrared Array Camera. Principal Investigator-led teams provided the instruments, while Ball Aerospace provided the remainder of the CTA. Lockheed Martin provided the solar panel and spacecraft bus. The observatory is approximately 4.5 m tall and 2.1 m in diameter; the mass at launch was 861 kg. The dust cover atop the CTA was jettisoned 5 days after launch. (Image courtesy of Ball Aerospace**)**

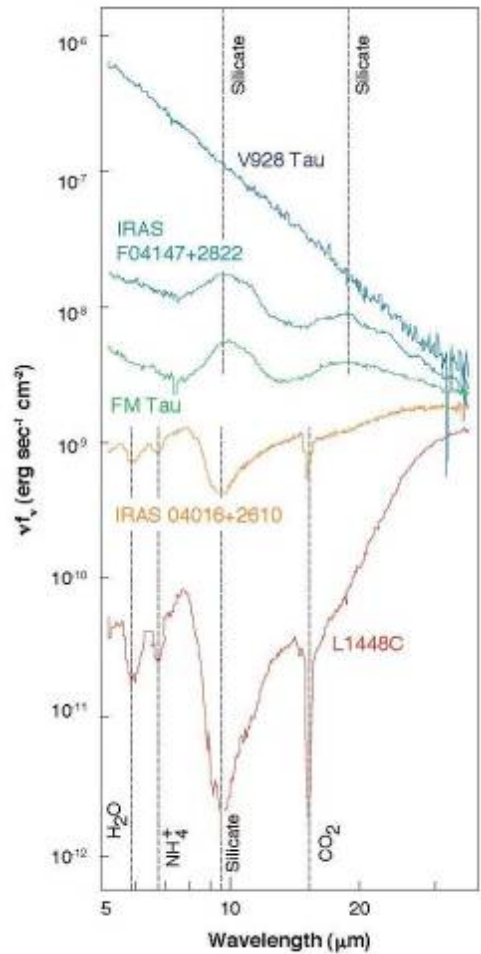

**Figure 2:** *Spitzer*-Infrared Spectrograph spectra of exemplars of the four classes of young stellar objects: the Class 0 object L1448C (Najita et al., in preparation the Class I object IRAS 04016+2610 (Watson et al. 2004); two Class II objects with substantially different small-dust composition, FM Tau and IRAS F04147+2822 (E. Furlan, L. Hartmann, N. Calvet, P. D'Alessio, R. Franco-Hernandez, et al., submitted; and D.M. Watson, J.D. Leisenring, E. Furlan, C.J. Bohac, B. Sargent, et al., submitted); and the binary Class III object V928 Tau (E. Furlan, L. Hartmann, N. Calvet, P. D'Alessio, R. Franco-Hernandez, et al., submitted). Identifications are shown for the most prominent spectral features, which are due to ices and minerals.  The spectra of FM Tau, IRAS F04147, and V928 Tau have been offset by factors of 50, 200, and $10^4$, respectively.

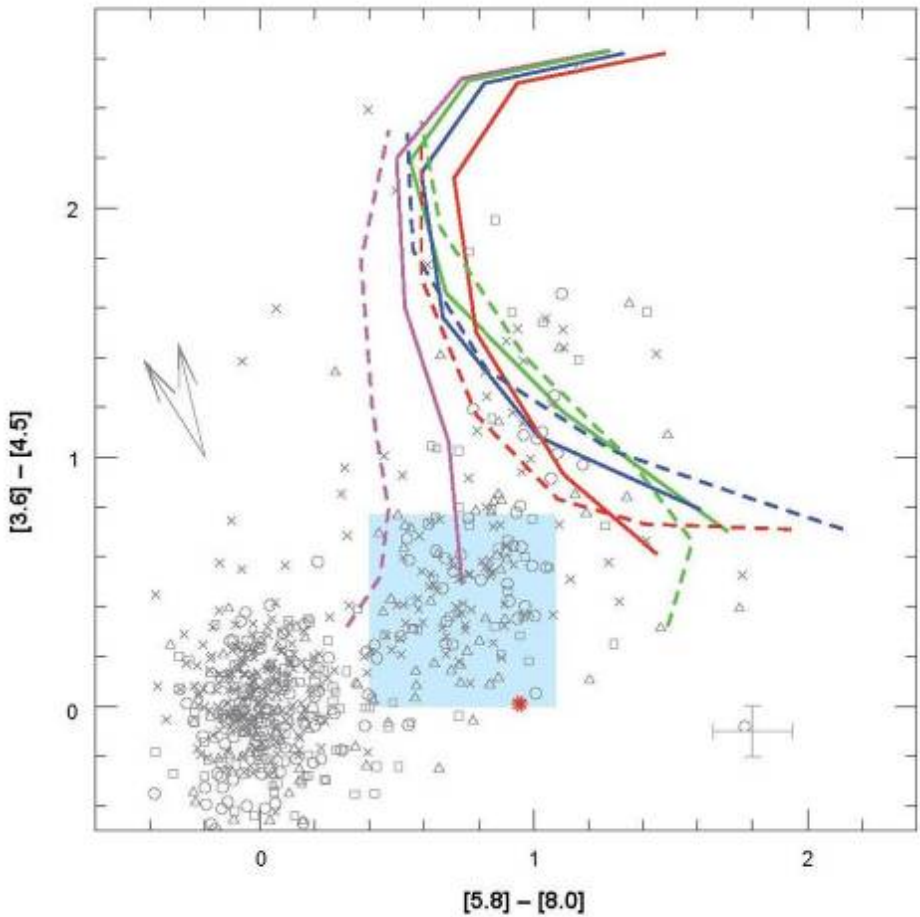

**Figure 3**:  Model colors and measured Infrared Array Camera (IRAC) colors for four young stellar clusters—S140 (*squares*), S171 (*circles*), NGC 7129 (*triangles*), and Cep C (*crosses*) from Allen et al. (2004). Representative error bars are shown. The data fall into three main groups: a group at (0,0) that contains background and foreground stars and Class III objects with no infrared excess in the IRAC bands, a group that occupies the Class II region (within the *blue box*), and a group that lies along the Class I theoretical locus with luminosity $L > 1$ $L_\odot$. The identification of the Class I and II sources is based on the agreement of the colors with the distribution of model colors (Kenyon, Calvet & Hartmann 1993; Calvet et al. 1994; D'Alessio et al. 1998, 1999, 2001, 2005b; D'Alessio, Calvet & Hartmann 2001 Extinction vectors are shown for $A_v$ = 30 mag, using the two extremes of the six vectors calculated by Megeath et al. (2004). (Reproduced by permission of the Am. Astron. Soc.)

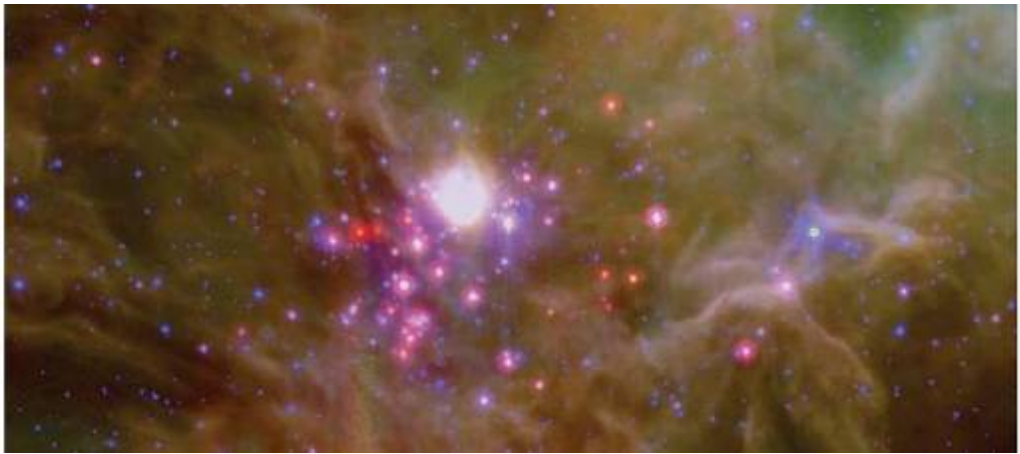

**<u>Figure 4</u>:**  Image of the Spokes cluster in NGC 2264. Color coding: blue = IRAC 3.6 $\mu$m; green = IRAC 8.0 $\mu$m; red = MIPS 24 $\mu$m. The region shown is about 5 × 10 arcmin in extent. North is up, and east is to the left. The image shows unusual linear alignments of the brightest 24-$\mu$m sources; Teixeira et al. (2006) show that they preferentially have a separation of 20″, close to the Jeans length estimated for the parent cloud of the cluster given the adopted distance of 800 pc

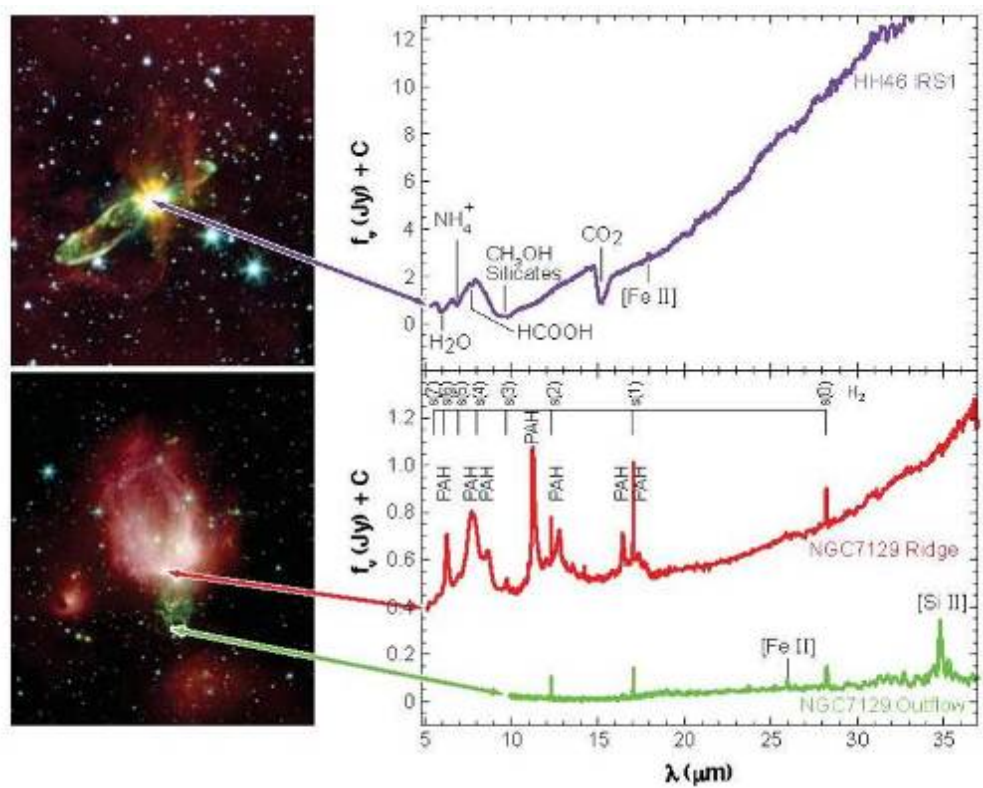

**Figure 5:** Outflows, nebulae, and embedded sources. Lower left: Infrared Array Camera (IRAC) image of the NGC 7129 region, ~10′ on a side. The distance to NGC 7129 is ~1.25 kpc. The NGC 7129 image is rotated so that the vertical direction is southwest on the sky; the HH46-47 image is rotated so that south is up and east is to the right. Color coding (for both left images): blue = 3.6 μm, green = 4.5 μm, and red = 8.0 μm. Lower right: Infrared Spectrograph (IRS) spectra obtained over ~20″ × 20″ regions at the indicated positions of the NGC 7129 outflow (green) and the reflection nebulosity (red) (Morris et al. 2004). The outflow spectrum stops at the IRS high-resolution limit of 10 μm; Infrared Space Observatory (ISO) spectra (e.g., Rosenthal, Bertoldi & Drapatz 2000) show that H2 and CO emission lines from outflows cluster in IRAC band 2 (4–5 μm), giving the outflow its green color in this image. The reflection nebulosity looks red because of strong PAH emission at 8 μm. Upper left: IRAC image of the HH46-47 system, ~6′ on a side. Upper right: Spectrum of the bright central source embedded between the lobes in HH46-47 (Noriega-Crespo et al. 2004). The deep absorption features arise in the torus that confines the outflow. See van Dishoeck (2004) for similar results from ISO. (Right image reproduced by permission of the Am. Astron. Soc. Top left image credits: NASA/JPL-Caltech/A. Noriega-Crespo [SSC/Caltech], Digital Sky Survey. Bottom left image credit: NASA/JPL-Caltech/T. Megeath [Harvard-Smithsonian Cent. for Astrophys.])

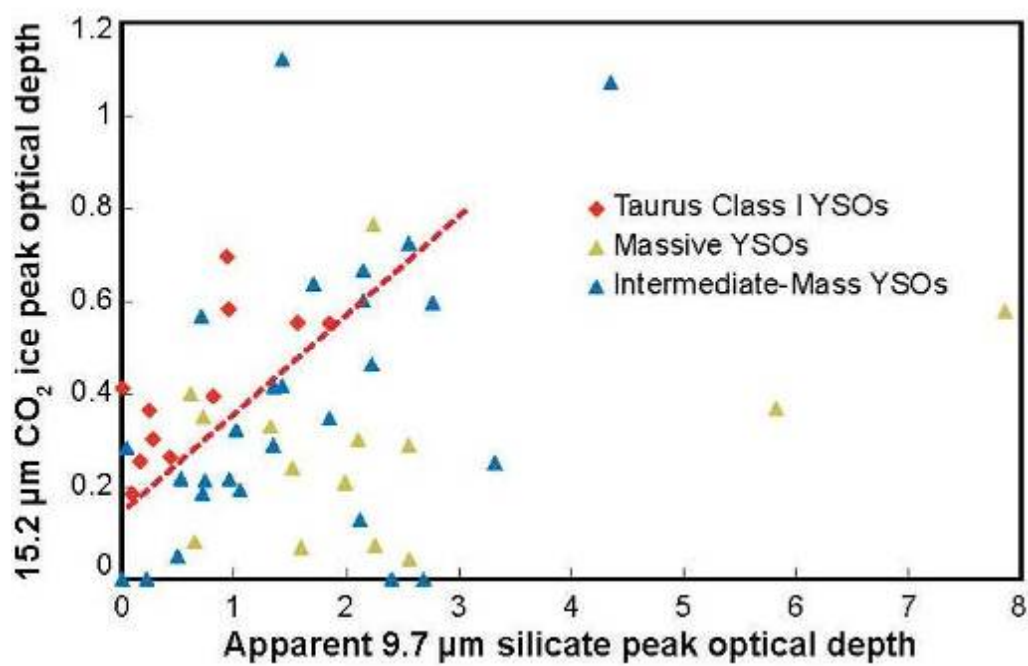

**Figure 6:** Peak optical depth in the 15.2-μm feature of CO2 ice, plotted as a function of the apparent (i.e., excess of absorption over emission) peak optical depth of the 10-μm silicate feature, for low-mass Class I objects in Taurus (F. Markwick-Kemper, in preparation; see also Watson et al. 2004) observed by Spitzer, and toward intermediate- and high-mass YSOs observed by the Infrared Space Observatory (Alexander et al. 2003, Gibb et al. 2004).

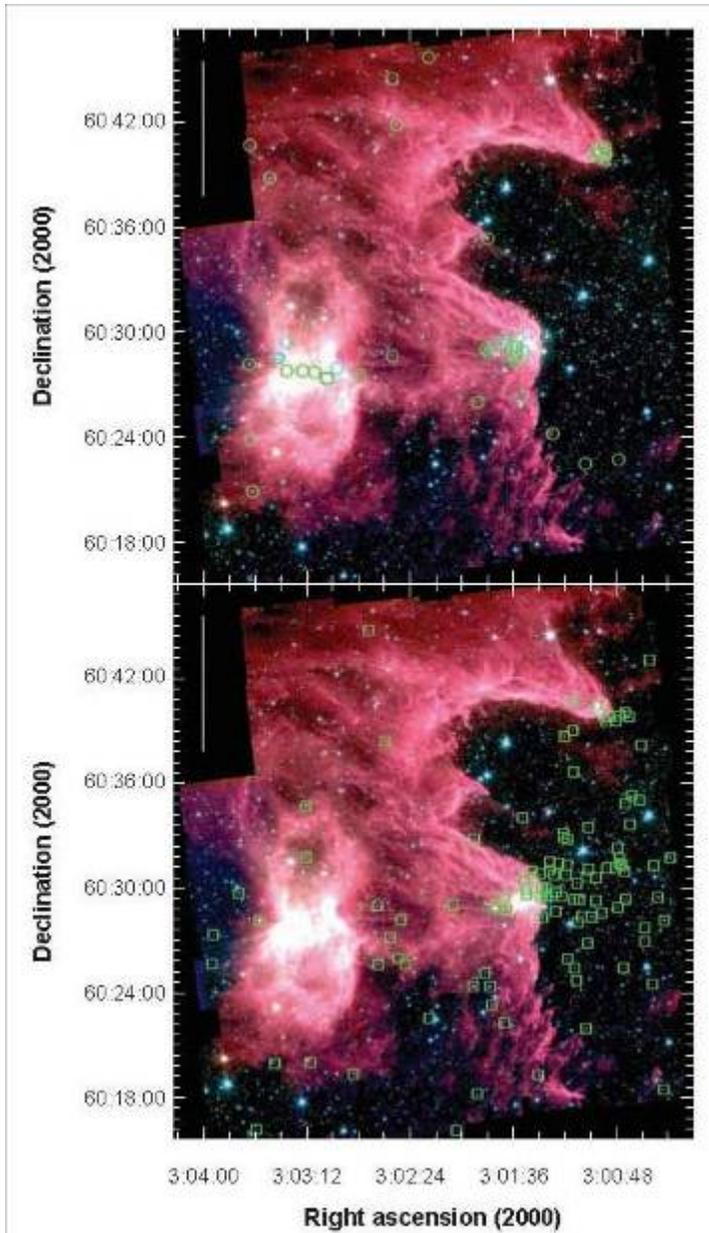

**Figure 7:** Infrared Array Camera (IRAC) images of the GL4029 region, which lies at a distance of about 2 kpc. A bright star just to the right of the region imaged sculpted the complex structure shown out of a molecular cloud. In the top image, circles show the positions of Class I objects identified by their IRAC colors; similarly, squares in the bottom image show the positions of Class II objects, which are much more widely distributed. Color coding by IRAC Band: blue = 3.6 $\mu$m, green = 4.5 $\mu$m, orange = 5.8 $\mu$m, and red = 8.0 $\mu$m. (Image credit: NASA/JPL-Caltech/L. Allen [Harvard-Smithsonian Cent. For Astrophys.])

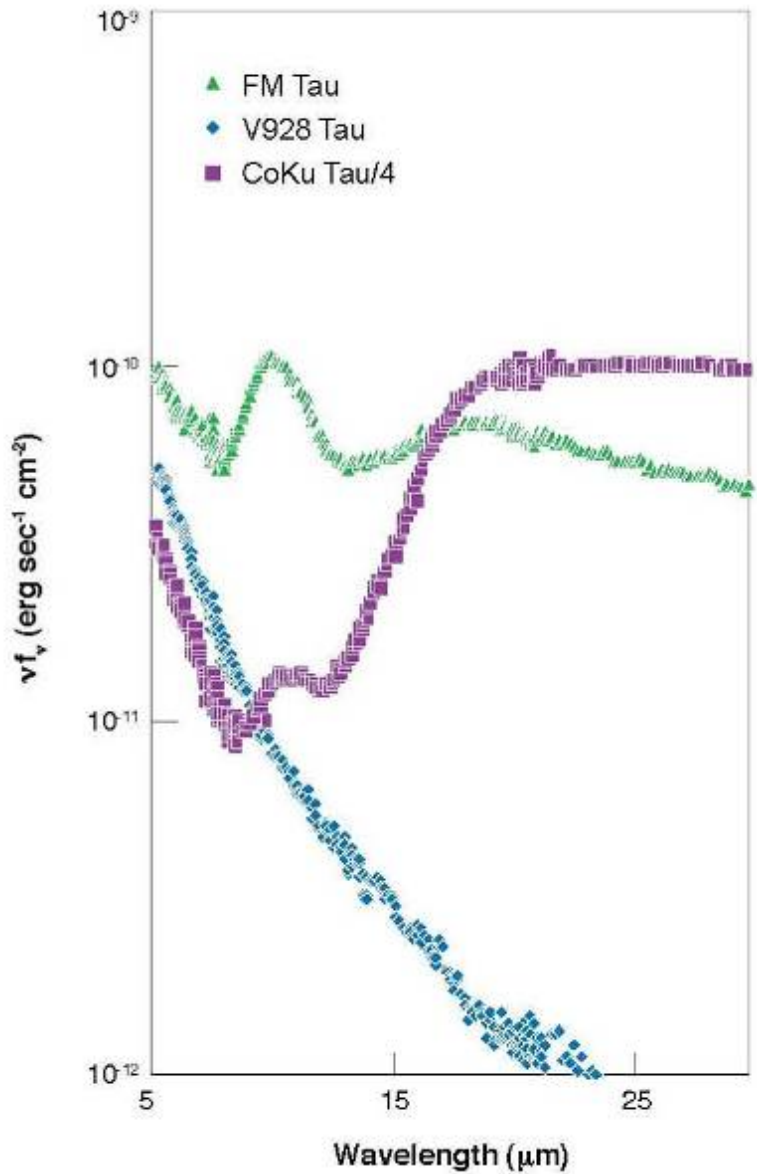

**Figure 8:** Infrared Spectrograph spectra of three Class II/III young stellar objects in Taurus. CoKu Tau/4 shows a flux deficit in the 5–15 $\mu$m region, indicative of a central clearing in its circumstellar disk. (Forrest et al. 2004, revised and reproduced by permission of the Am. Astron. Soc.). CoKu Tau/4 is an example of a transition disk, as discussed in Section 3.2.2

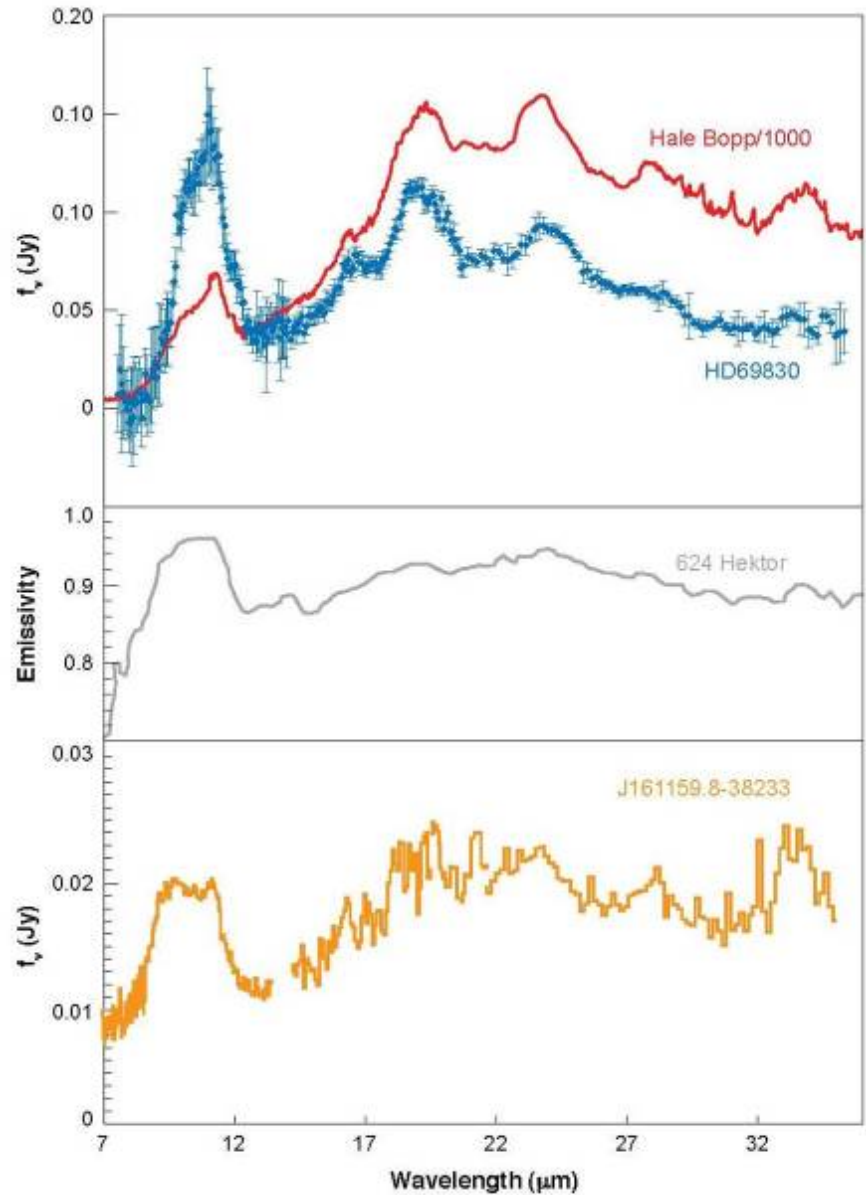

**Figure 9:** Infrared Spectrograph low-resolution spectra of the 2-Gyr-old K star HD69830 (top; Beichman et al. 2005a); the Trojan asteroid 624 Hektor (emissivity spectrum, middle; J.P. Emery, D.P. Cruikshank & J. Van Cleve, submitted); and the candidate brown dwarf J161159.8-38233 in Lupus (bottom; B. Merin, et al., in preparation are compared with the spectrum of comet Hale Bopp (top; Crovisier et al. 1996). All sources show the characteristic emission features of Mg-rich crystalline silicates. (Top spectra reproduced by permission of the Am. Astron. Soc. Middle spectrum reprinted from Icarus with permission from Elsevier.)

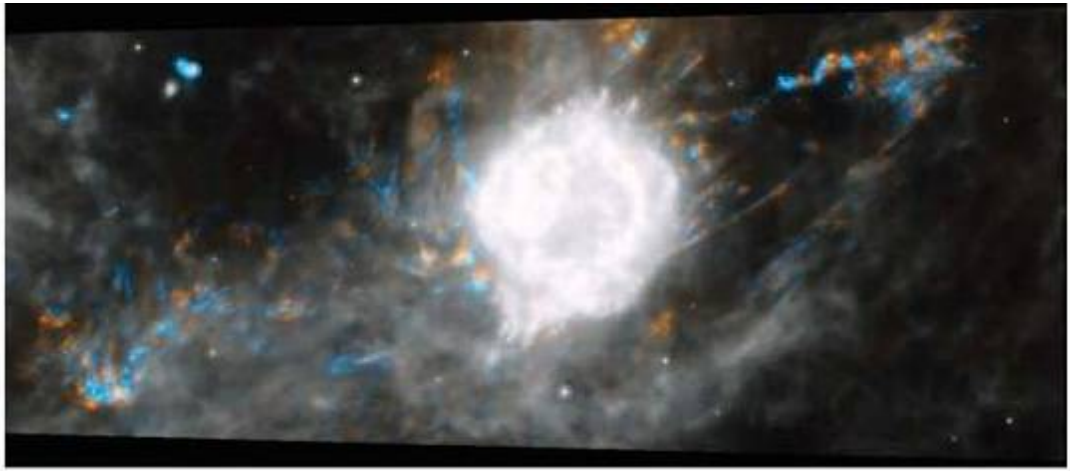

**Figure 10:** Two images of the Cas A supernova remnant and its environs at 24 μm, taken one year apart, superposed to show the propagation of a light echo (Krause et al. 2005). The blue image dates from November 2003, whereas the orange image was taken in December 2004. The strip shown is approximately 20 arcmin in length, oriented with east down and north to the left. [Image credit: NASA/JPL-Caltech/O. Krause (Steward Observatory)]

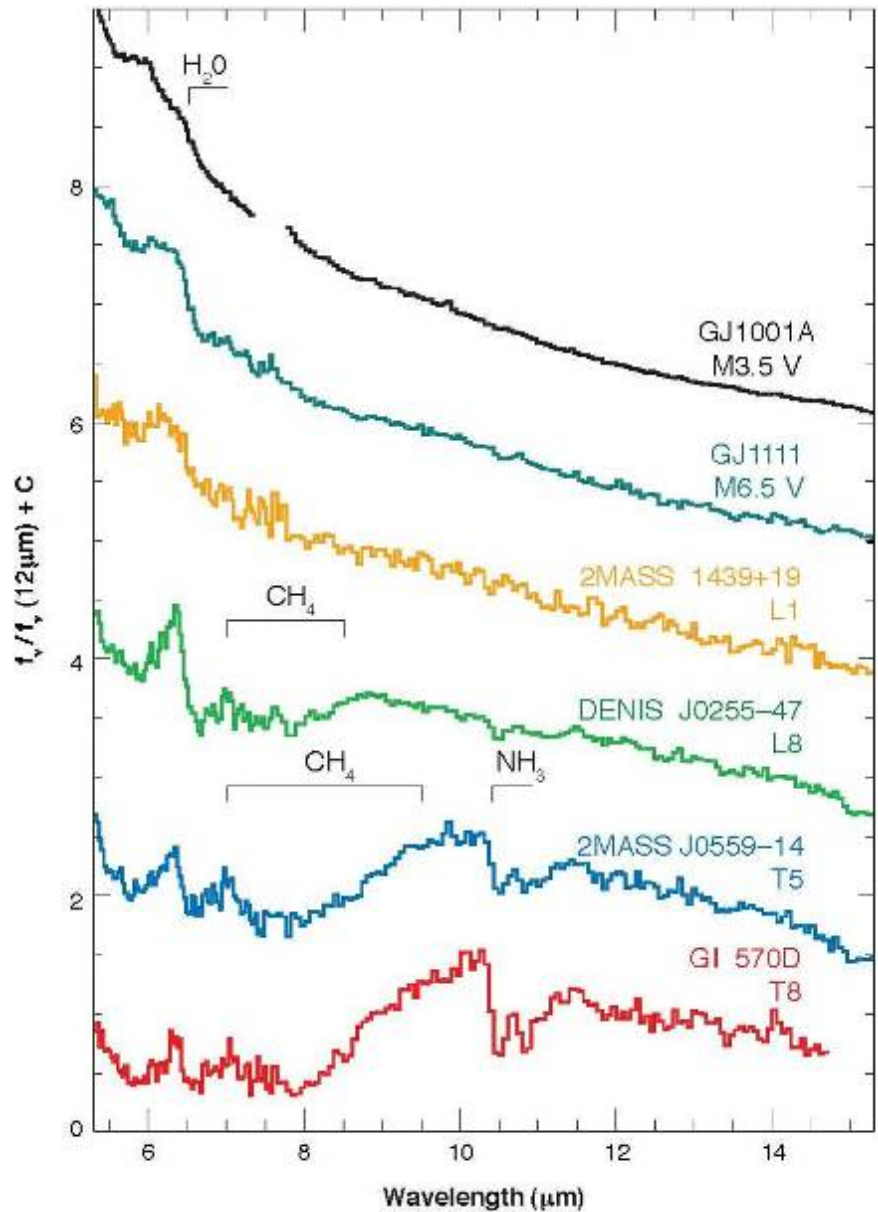

**Figure 11:** A selection of M, L, and T brown dwarf spectra taken with the *Spitzer* Infrared Spectrograph. The effective temperatures of these objects range from approximately 3400 K for the M3.5V object down to 800 K for the T8 dwarf. The molecular species responsible for the obvious features are indicated. The smaller-scale spectral features are also due to water, methane, and ammonia, with their relative importance depending on the temperature. Because there are no observed temperature inversions in the atmospheres of these objects, all of the features are seen in absorption (Cushing et al. 2006).

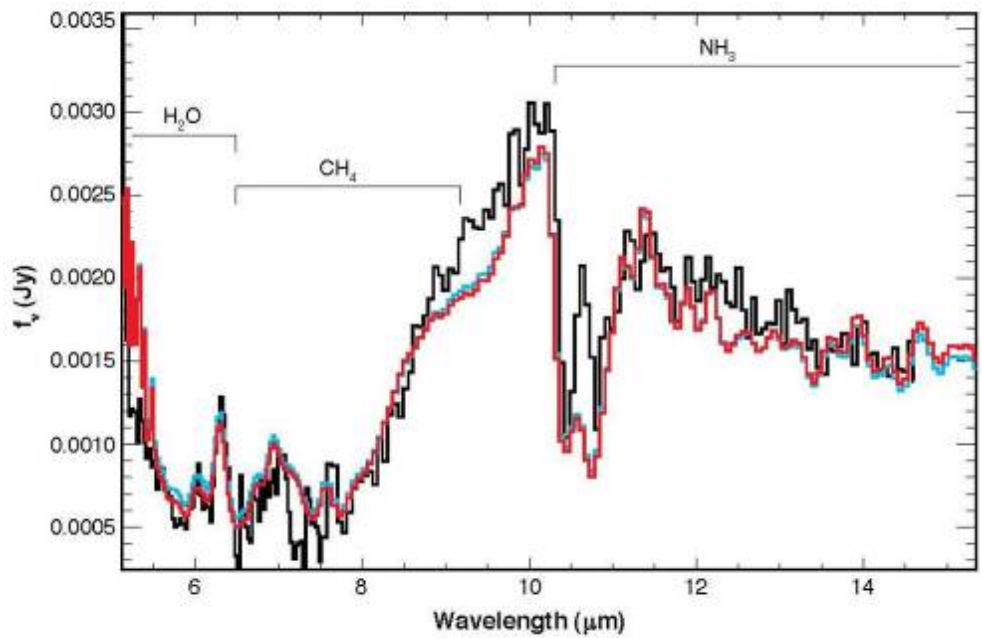

**Figure 12:** The observed spectrum of the T8 dwarf Gl570D, plotted in black together with two model spectra. The blue and red lines indicate models with and without clouds, respectively. Although the cloud model predicts that the clouds will lie below the photosphere in such a cool object (resulting in near-perfect agreement between the red and blue lines), the observed spectrum is brighter than expected in the 9.8-μm region near the silicate feature (Cushing et al. 2006). The models compared to the data are taken from Marley et al. (2002, 2003) and Saumon et al. (2003).

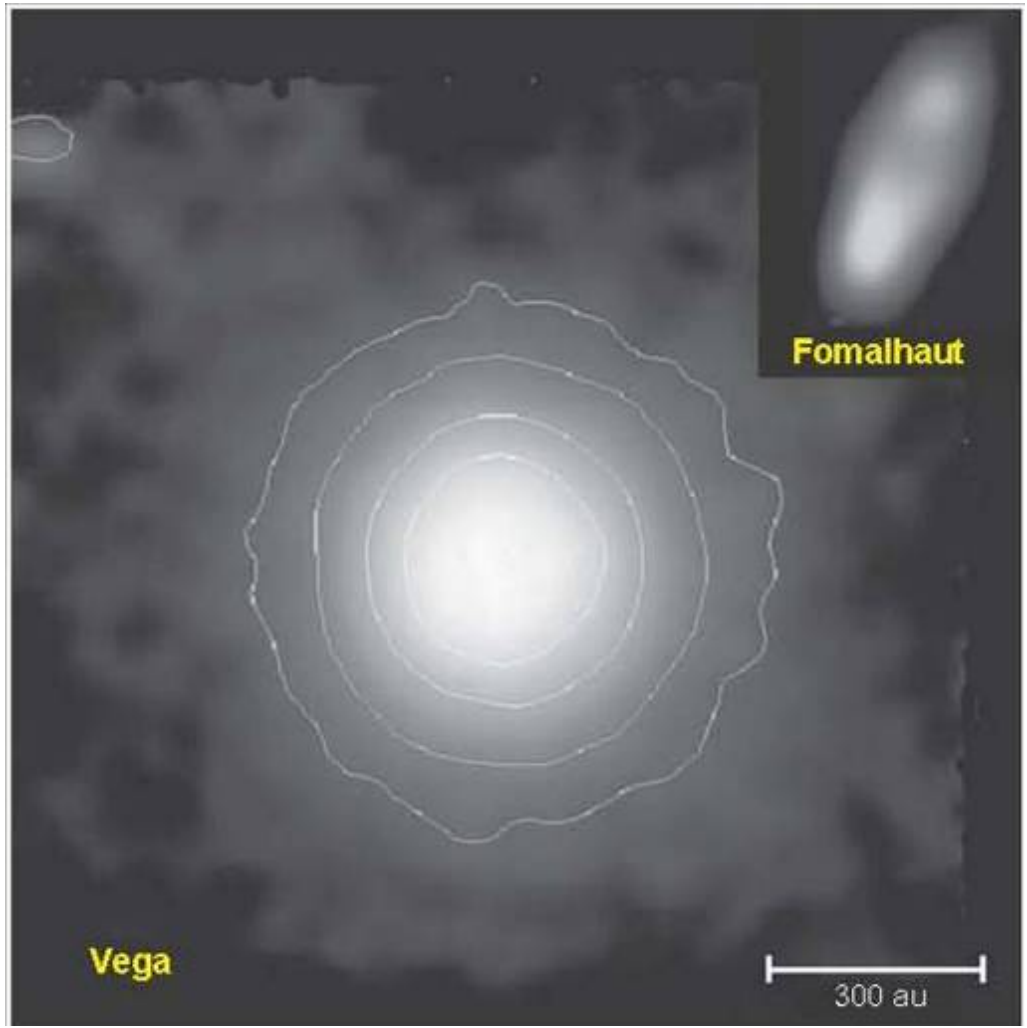

**Figure 13:** The 70 μm Multiband Imaging Photometer for Spitzer (MIPS) image of the debris disk around Fomalhaut (Stapelfeldt et al. 2004) is shown on the same spatial scale as that around Vega (Su et al. 2005) to illustrate the great variety in appearance of debris systems with similar spectral energy distributions. A logarithmic display is used for Vega to highlight the great extent of the dust orbiting this star; by contrast, the smaller Fomalhaut disk has a sharp outer edge. (Reproduced by permission of the Am. Astron. Soc.)

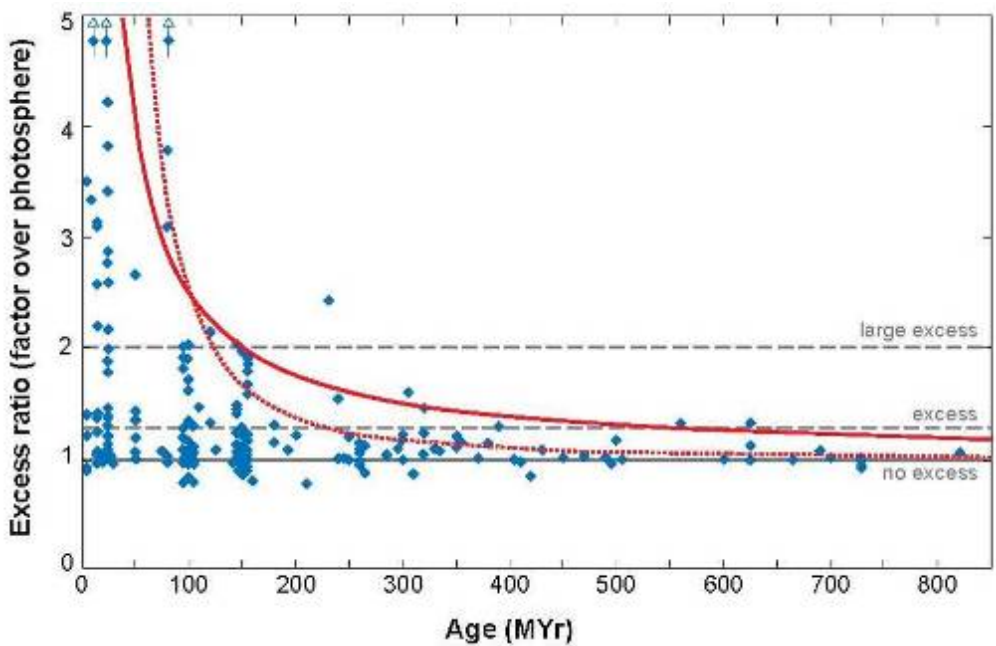

**Figure 14:** 24-μm excess versus age for a large sample of A stars (Rieke et al. 2005). Excess emission is indicated as the ratio of the measured flux density to that expected from the stellar photosphere alone: a value of 1 represents no excess (solid gray line). Additional dashed gray horizontal lines show the threshold for detection of an excess (1.25) and the one for a large excess (2). The upward-pointing arrows are, from left to right, HR 4796A, βPic, and HD 21362. The thin solid line is an inverse time dependence, whereas the thin dashed line is inverse time squared. Age uncertainties are roughly a factor of 1.5 below 200 Myr (where ages are almost entirely from cluster and moving group membership) and roughly a factor of 1.5–2 above 200 Myr (where many ages are assigned by placing the stars on the Hertzsprung-Russell diagram). (Reproduced by permission of the Am. Astron. Soc.)

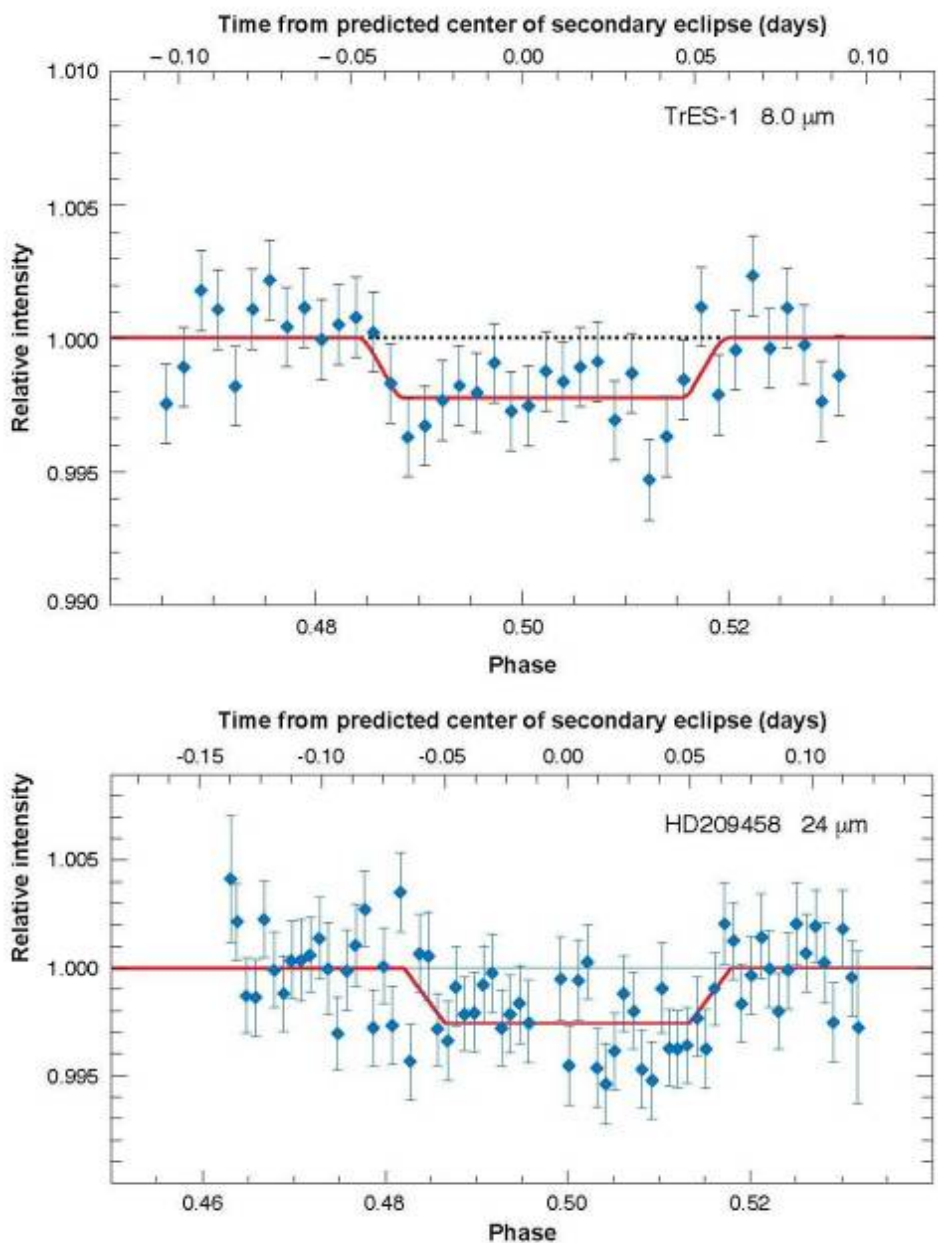

**Figure 15:** Charbonneau et al. (2005) and Deming et al. (2005) detected light from secondary eclipses of, respectively, the TrES-1 (top) and HD209458 (bottom) systems. The drop in relative intensity in each plot coincides with the disappearance of the planet behind the star. Solid red lines show best-fit secondary-eclipse models (Charbonneau et al. 2005, revised and reproduced by permission of the Am. Astron. Soc. Deming et al. 2005 adapted by permission from Macmillan Publishers Ltd: Nature, copyright 2006.)

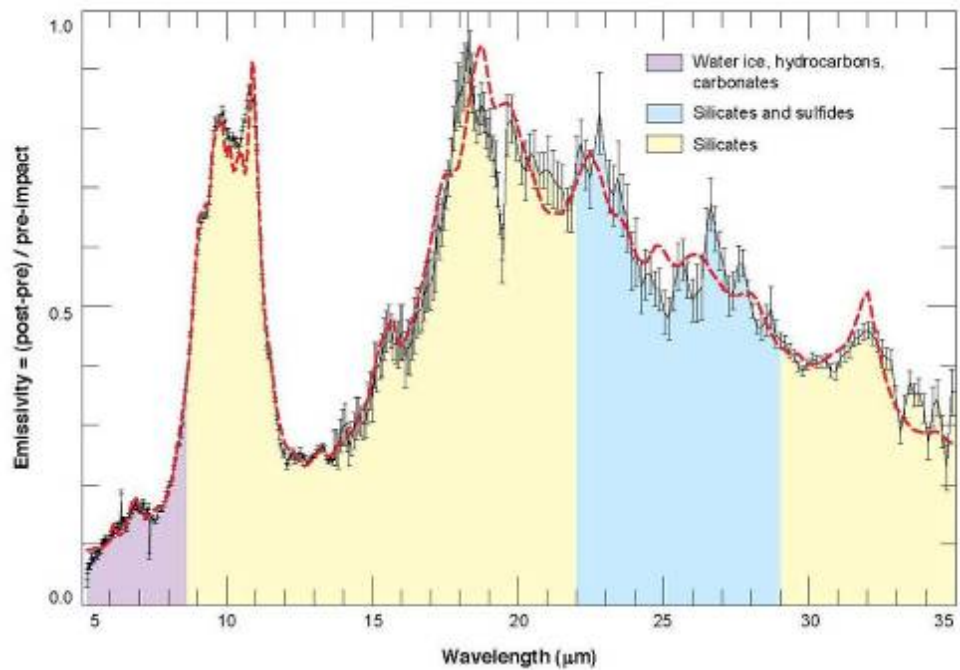

**Figure 16:** The data points show the emissivity spectrum of the ejecta from comet Tempel 1 as measured 45 minutes after the Deep Impact encounter (Lisse et al. 2006). The red line is the best fit model. The color coding indicates the dominant materials in each wavelength interval.